\documentstyle[12pt,aaspp4]{article}

\def\<{{\langle}}
\def\>{{\rangle}}
\def\cm{{\rm cm}}
\def\G{{\rm G}}
\def\K{{\rm K}}
\def\yr{{\rm yr}}
\def\Myr{{\rm Myr}}
\def\msun{{\rm\,M_\odot}}
\def\lsun{{\rm\,L_\odot}}
\def\kms{{\rm km\, s^{-1}}}
\def\pc{{\rm pc}}
\def\va{v_{\rm A}}

\def\cs{c_{\rm s}}
\def\vt{v_{\rm turb}}
\def\ts{t_{\rm s}}
\def\tc{t_{\rm c}}
\def\ta{t_{\rm A}}
\def\Lj{L_{\rm J}}
\def\nj{n_{\rm J}}
\def\Ew{E_{\rm W}}
\def\Eb{E_{\rm B}}
\def\Ek{E_{\rm K}}
\def\Pw{P_{\rm wave}}
\def\alf{Alfv\'en }
\def\bB{{\bf B}}
\def\nh2{n_{H_2}}
\def\cf{cf.\ }

\def\fbeta{{c_s^2/v_{{\rm A}}^2}}
\def\bv{{\bf v}}
\def\del{\partial}
\def\rb{{\bar{\rho}}}
\def\au{{\rm\,AU}}

\lefthead{Gammie and Ostriker}
\righthead{Nonlinear Alfv\'en Waves}

\begin{document}

\title{Can Nonlinear Hydromagnetic Waves Support a 
Self-Gravitating Cloud?}

\author{Charles F. Gammie and Eve C. Ostriker}
\affil{Harvard-Smithsonian Center for Astrophysics, MS-51 \\
60 Garden St., Cambridge, MA 02138}

\begin{abstract}

Using self-consistent magnetohydrodynamic (MHD) simulations, we
explore the hypothesis that nonlinear MHD waves dominate the internal
dynamics of galactic molecular clouds.  Our models employ an
isothermal equation of state and allow for self-gravity.  We adopt
``slab-symmetry,'' which permits motions $\bv_\perp$ and fields
$\bB_\perp$ perpendicular to the mean field, but permits gradients only
parallel to the mean field.  This is the simplest possible geometry
that relies on waves to inhibit gravitational collapse along the mean field.
In our simulations, the \alf
speed $\va$ exceeds the sound speed $\cs$ by a factor $3-30$, which is
realistic for molecular clouds.  We simulate the free decay of a
spectrum of \alf waves, with and without self-gravity.  We also
perform simulations with and without self-gravity that include
small-scale stochastic forcing, meant to model the mechanical energy
input from stellar outflows.

Our major results are as follows: 
(1) We confirm that the pressure associated with fluctuating transverse 
fields can inhibit the mean-field
collapse of clouds that are unstable by Jeans's criterion. 
Cloud support requires the energy in \alf-like disturbances
to remain comparable to the cloud's gravitational binding energy.  
(2) We characterize the turbulent energy spectrum and density structure in
magnetically-dominated clouds.  The perturbed magnetic and transverse
kinetic energies are nearly in equipartition and far exceed the
longitudinal kinetic energy.
The turbulent spectrum evolves to a power-law shape, approximately
$v_{\perp,\, k}^2\approx B_{\perp,\, k}^2/4\pi\rho\propto k^{-s}$ with 
$s\sim 2$, i.e. 
approximately consistent with a
``linewidth-size'' relation $\sigma_v(R)\propto R^{1/2}$.  The
simulations show large density contrasts, with high density regions
confined in part by the pressure of the fluctuating magnetic field.
(3) We evaluate the input power required to offset dissipation through
shocks, as a function of $\cs/\va$, the velocity dispersion
$\sigma_v$, and the characteristic scale $\lambda$ of the forcing.  In
equilibrium, the volume dissipation rate is $5.5
(\cs/\va)^{1/2}(\lambda/L)^{-1/2}\times \rho \sigma_v^3/L$, for a
cloud of linear size $L$ and density $\rho$.  (4) Somewhat
speculatively, we apply our results to a ``typical'' molecular cloud.
The mechanical power input required for equilibrium (tens of $\lsun$),
and the implied star formation efficiency ($\sim 1\%$), are in rough
agreement with observations.
Because this study is limited to slab symmetry and excludes ion-neutral
friction, the dissipation rate we calculate probably provides
a lower limit on the true value.  

\end{abstract}


\section{Introduction}

The internal dynamics of star-forming galactic molecular clouds is not
yet understood.  Two central questions are (1) what prevents the
clouds and their subcomponents from collapsing under their own
weight; and (2) what generates and controls the turbulent fluid
velocities that broaden molecular lines far beyond the thermal speed
$\cs$ (e.g. \cite{shu87}).  One model which has been proposed (e.~g.
\cite{sca82}) is that the clouds are comprised of clumps on essentially 
ballistic, collisionless orbits.  However, while clouds are observed to be 
clumpy, the volume filling
factor of clumps in the clouds $f\sim 0.03-0.08$ (e.g. \cite{per85};
\cite{wil95}) implies a clump-clump collision time $t_{\rm
collis}<(4/3){R_{\rm clump}/(f v_{\rm clump})} \sim 10^7 \yr$, which
makes the clouds at most marginally collisionless over their lifetimes
(\cite{bli80}). The clumps are not themselves thermally supported, and
they appear to have larger internal filling factors and smaller ratios
of internal collision time to dynamical time.  Although internal
velocities may be generated by a cloud's self-gravity, purely
hydrodynamic turbulence -- either clumpy or smooth -- cannot in itself
support a structure for longer than the effective collision time
(equal to the eddy-turnover time for a uniform fluid) because it would
dissipate in shocks (see \cite{elm85} and references therein).  The
orbiting-clump model therefore probably cannot account for the
internal dynamics of molecular clouds at all scales.
\footnote{In particular, it is not apparent how a collisionless system
could generate non-self-gravitating clumps whose internal velocity
dispersions scale with their size (\cf \cite{ber92}).} Rather than
assuming a clumpy mass distribution {\it a priori}, it seems better to
start with a full fluid model with a compressible equation of state,
so that clumping can be treated self-consistently.  Such a model must
have some internal stress far more powerful than gas pressure in order
to control supersonic motions.

For some time, magnetic fields have been considered the leading
candidate for mediating clouds' internal motions and counteracting
gravity (see the recent reviews of \cite{shu87};  \cite{mck93}).
Magnetic processes have also been identified as likely instruments for
generating density structure within clouds (e.g. \cite{elm90};
\cite{elm91}), which is observed at all scales down to the limiting
telescopic resolution (\cite{fal91}; \cite{fal92}).  Measured field
strengths $B_\parallel$ based on OH Zeeman splittings are in the range
$10-30 \mu \G$ (\cite{cru93}) for the line-of-sight field in
moderate-density regions $\nh2\gtrsim 1000 \cm^{-3}$ (for random
orientations the mean total field strength is twice as large).  Fits
incorporating additional data from weak-field, low-density HI Zeeman
splitting and strong-field, high-density OH maser Zeeman splitting
yield $B\approx 1.5 (\nh2/1 \cm^{-3})^{1/2} \mu \G$ (\cite{hei93},
and references therein).  Based on these data, the magnetic field has
an energy density comparable to the kinetic (and gravitational) energy
densities, and therefore can be dynamically important.  More
specifically, \cite{mye88a} show that magnetic, kinetic, and
gravitational energies are comparable in detail for several clouds at
a range of scales, suggesting virial equilibrium.  The field topology
within molecular clouds remains uncertain.  In optical wavelengths,
the linear polarization directions of background stars shining through
low-density regions undulate smoothly across cloud complexes (e.g.
\cite{mon84}).  To trace higher-density gas within clouds, longer
wavelengths are needed.  Maps of polarized $100\mu$ thermal emission in
several high-mass star-forming regions 
((\cite{dot95}), \cite{hil95}, \cite{dav95}) also show orderly variation
across the cloud.  If in both cases the polarization is caused by
field-aligned dust grains, the data imply smoothly-varying mean
fields.  These preliminary indications on field geometry, if
confirmed, permit a conceptual separation into cloud support
perpendicular to, and parallel to, a slowly-varying, untangled, mean field.

To date, most theoretical work on magnetic fields in star-forming
regions has concentrated on the role of smooth fields in quasi-static
equilibria or configurations undergoing laminar rotation and/or
collapse (see the reviews of \cite{nak84}; \cite{mou91};
\cite{mck93}).  The absence of turbulent velocities $\vt$ exceeding
$\cs$ in the small, dense cloud cores observed to be the sites of
low-mass star formation (see, e.g. \cite{ful92}) makes them amenable
to quasistatic theories.  To the extent that turbulent magnetic and
Reynolds stresses can be included via a barotropic pressure, such
calculations can also be applied to cases where $\vt>\cs$.
Axisymmetric calculations of field-frozen equilibria have quantified
the importance of field support perpendicular to the mean field
direction, which can be expressed succinctly in terms of the
mass-to-magnetic flux ratio, $M/\Phi$ (\cite{mou76a}; \cite{mou76b};
\cite{tom88}). The value of this evolutionary invariant determines
whether or not an equilibrium can be sustained.

While static or time-averaged fields are likely key to cloud support at
both small and large scales, they do not oppose gravity in the mean
field direction, and by definition cannot produce a large velocity
dispersion.  For clumps within clouds (reviewed by \cite{bli93}; see
also \cite{ber92}), and massive cloud cores (e.g. \cite{cas95}),
however, molecular line observations exhibit linewidths in excess of
$\cs$.  The inferred hypersonic bulk velocities were attributed to MHD
waves shortly after their discovery (\cite{aro75}).  For Alfv\'en
waves, the fluctuating component of the field provides a pressure that
acts along the mean field, and can therefore oppose gravity in that
direction (\cite{dew70} ;  \cite{shu87}; \cite{fat93}; \cite{mck95}).

The theory of \cite{dew70} calculates the influence of small-amplitude
MHD waves on the background state of the fluid, using a
locally-averaged Lagrangian.  For \alf waves, the effect of the waves
is expressed through an isotropic wave pressure $P_{\rm wave}=\<
|\delta {\bf B}|^2\> /8\pi$, where the magnetic disturbance is $\delta
{\bf B}$.  Recently, \cite{mck95} have used Dewar's theory to show that
small-amplitude Alfv\'en waves propagating along a density gradient
obey $P_{\rm wave}\propto \rho^{1/2}$ (implying effective polytropic
index $\gamma_p=1/2$), while waves trapped in a contracting cloud obey
$P_{\rm wave}\propto \rho^{3/2}$ (implying effective adiabatic index
$\gamma_a=3/2$).  Since gas spheres are dynamically stable to
adiabatic perturbations when $\gamma_a > 4/3$ (e.g. \cite{cox80}),
large amplitude \alf waves could potentially support a cloud against
collapse if they suffered minimal decay, or loss (cf. \cite{elm85})
into the surrounding medium, and obeyed the same scaling.  A crucial
unknown is the decay rate of arbitrary amplitude MHD waves in
conditions appropriate for a molecular cloud.  If Alfv\'en waves are
responsible for the internal linewidths of molecular clouds and support
against gravity and external pressure, then any decay must be
replenished if a quasi-equilibrium is to be maintained.  Prior to
high-mass star formation, the ultimate source of new wave energy must
be the gravitational potential of the cloud (\cf \cite{mes56};
\cite{nor80}; \cite{fal86}; \cite{mck89}).   Potential energy is liberated 
by overall cloud contraction and/or star formation (with winds);  for 
quasi-equilibrium the respective rates would depend on how
fast waves decay.  At present, estimates of decay rates rely on
analytic calculations of linear damping by ion-neutral friction,
nonlinear single-wave steepening (see \cite{zwe83} and references
therein), or simply dimensional analysis in terms of the internal
velocity dispersion and size (see \cite{bla87} and references
therein).

Previous theoretical studies of MHD waves in molecular clouds have
concentrated on the linear, adiabatic, WKB limit in analyzing wave
dynamics (\cite{fat93}; \cite{mck95}; \cite{zwe95}).  This approach is
accurate when the amplitude of the wave is small, when the wavelength
of the waves is much smaller than the size of the system, and when the
wave damping is weak.  Since the first two of these assumptions do not
strictly apply in molecular clouds, and the third may not either, it is
interesting to see if we can relax them somewhat.   In this paper, we
undertake to study the full nonlinear development of moderate-amplitude
MHD disturbances in a self-gravitating system, using numerical
simulations.  We concentrate our attention on the most basic questions
of how well clouds can be supported against gravity by nonlinear
disturbances, and how long it takes MHD turbulence to decay.  By
employing simulations, we can generalize ideas of support by simple
Alfv\'en waves to include arbitrary self-consistent disturbances in the
magnetic field, fluid flow, and density structure.  We can test how far
the linear-theory predictions for wave support carry over to the
nonlinear regime, and go beyond the purview of simple linear theory to
investigate the growth of structure, cascade of energy between scales,
and associated dissipation.

Expedience demands some sacrifices of realism for this first study.
Our most severe simplification is to restrict the motions to
plane-parallel geometry, so all dynamical variables are functions of
one independent space variable and time.  Thus we allow for transverse
motions but not transverse derivatives.  Another idealization which is
less serious for the large-scale motions we consider is the neglect of
ambipolar diffusion.  We shall also assume the gas is isothermal.

The plan of this paper is as follows.  In \S 2  we
review basic observational results for molecular clouds and the implied
timescales in the context of simple theoretical considerations of cloud
stability.  We then present the equations we shall solve, our
numerical method, and various tests used to verify its performance (\S
3).  In \S 4 we describe the series of simulations we have performed.
Finally, we apply our results to astronomical systems, discuss
directions for future research, and summarize our conclusions (\S 5).

\section{Summary of Observations and Theory}

\subsection{Cloud Properties and Scalings}\label{S: cloud properties}

At present, the overall dynamical characterization of molecular clouds
(e.g. delayed collapse, quasistatic contracting or expanding
equilibrium) remains uncertain.  In particular, the uncertainties
about formation mechanisms and difficulty of assigning ages hampers
efforts to deduce evolutionary trends (e.g. \cite{elm93}).  Arguments
about the dynamical state are therefore indirect.  The presence of
several-million-year-old stars within cloud complexes indicates they
have survived long enough to exceed the characteristic gravitational
collapse times of at least the (higher-density) clumps; yet, a general
internal collapse has not occurred.  While cloud complex ages may not
exceed the collapse times at their (lower) mean density, they are
likely stabilized by something stronger than gas pressure because
wholesale collapse on the dynamical times would lead to an unacceptably
large galactic star formation rate (see review of \cite{shu87}).
Indeed, the presence of substructure at all scales within molecular
clouds reveals that something much more interesting than a
nearly-pressure-free collapse is taking place.

Spectral observations of molecular transitions show hypersonic widths
everywhere except in the densest regions. These broad, often
non-Gaussian molecular lines ( \cite{fal90}; \cite{mie95}), give
evidence that the gas they trace is in a highly turbulent state, with
the turbulent amplitudes dependent on spatial scale.  For cloud
complexes, and massive component clumps and cores within them, internal
linewidths $\sigma_v$ typically increase with size scale $R$ as 
$\sigma_v(R)\propto R^{1/2}$ (this
type of scaling was first pointed out by \cite{lar81}; see also
\cite{dam86}, \cite{sol87}, \cite{fal87}, \cite{mye88b}).  Together
with the typical observed scaling of mean density with size as
$n\propto R^{-1}$ (\cite{lar81}), this relation implies that internal
velocities within these structures are roughly virial, consistent with
a response to self-gravitational confinement.  The internal kinetic
energy densities in the lower-mass, non-self-gravitating clumps (e.g.
\cite{car87}, \cite{lor89a}, \cite{lor89b}, \cite{her91}, \cite{fal92},
\cite{wil95})
are instead comparable to the mean internal energy density of the
surrounding GMC, suggesting a balance of stresses at the interface
between the clump and an interclump medium -- ``pressure confinement''
(\cite{ber92}).  The internal velocity and density structure in clouds
points to the importance of magnetic fields;  however, the lack of
detailed information on field strengths and topology makes it difficult
to decide how close to equilibrium any cloud, component clump, or core
really is (see \cite{goo94}).  Below we outline cloud observed properties, 
define relevant
timescales necessary for scaling our numerical simulations, and discuss
the corresponding simple stability requirements assuming uniform
conditions.

The average linear dimension $L$ is typically $40 \pc$ for GMCs, $10\pc$ 
for dark cloud complexes, and up to a few pc for the component clumps within
these complexes that contain most of the mass.  The volume-averaged
density $n_{\rm H_2}$ is typically $25-100 \cm^{-3}$ in cloud
complexes, and $10^3 \cm^{-3}$ in clumps.  Kinetic temperatures
range from $10-50\K$.  This implies an isothermal sound speed
$c_s = \sqrt{k T/\mu} = 0.19-0.41 \kms$, where $\mu = 2.4 m_p$ 
(e.g. \cite{bli93}; \cite{cer91}).  
A convenient measure of the importance of magnetic fields is the parameter
$\beta$, 
\footnote{Our $\beta$ differs by a factor of $2$ from the usual 
plasma $\beta_p \equiv (gas\ pressure)/(magnetic\ pressure)$.}
defined by
\begin{equation}\label{betadef}
\beta \equiv {c_s^2\over{v_A^2}} = {c_s^2\over B^2/4\pi\rho} =
	0.021
	\left({T\over{10\K}} \right)
	\left({n_{H_2}\over{10^2\cm^{-3}}} \right)
	\left({B\over{10\mu{\rm G}}}\right)^{-2}.
\end{equation}
$\beta < 1$ implies a magnetically-dominated regime.
Generally $\beta$ ranges between $0.1$ and $0.001$ in molecular clouds,
reaching unity only in cloud cores.  Notice that under isothermal,
field-freezing conditions, $\beta$ decreases as a cloud contracts
homologously ($B \propto n^{2/3}$).  It is therefore natural to expect the
importance of magnetic forces to increase relative to gas pressure
gradients as an efficiently-radiating cloud condenses out of the
interstellar medium.

The sound wave crossing time over a scale $L$ is 
\begin{equation}\label{tsound}
\ts = {L\over{c_s}} = 
	53 \left({L\over 10 {\pc}}\right)\left({T\over{10\K}}\right)^{-1/2}
	{\rm M}\yr,
\end{equation}
a characteristic gravitational collapse time is
\begin{equation}\label{tcollapse}
\tc = \left(\pi\over G\rho\right)^{1/2} =
	9.9 
	\left({n_{H_2}\over{10^2\cm^{-3}}}\right)^{-1/2}
	{\rm M}\yr,
\end{equation}
and the \alf wave crossing time over a
distance $L$ along a uniform field $B_0$ is 
\begin{equation}\label{talfven}
\ta = {L\over{v_A}} \equiv {L\sqrt{4\pi\rb}\over{B_0)}} =
	7.6
	\left({L\over 10 {\pc}}\right)
	\left({n_{H_2}\over{10^2\cm^{-3}}} \right)^{1/2}
	\left({B_0\over{10\mu{\rm G}}}\right)^{-1}
	{\rm M}\yr.
\end{equation}

\subsection{Stability Requirements in a Uniform Medium}\label{S: equilibrium}

With the definition of collapse time in equation (\ref{tcollapse}),
the ratio of structure scale $L$ to the Jeans length 
$\Lj\equiv \cs (\pi/G\rho)^{1/2}$
satisfies $L/L_{\rm J}=\ts/\tc$, so structures with $\ts>\tc$ are above
the Jeans limit and cannot be supported by thermal pressure alone.
The importance of a mean field $\bB_0$ to cloud dynamics
can be measured by comparing the Alfv\'en crossing time $\ta=L/\va$ 
to the characteristic collapse time $\tc$,
\begin{equation}
{\ta\over\tc} = 0.76 \left({L\over 10\pc}\right) \left({\nh2\over 10^2
\cm^{-3}}\right) \left({B_0\over 10\mu {\rm G}}\right)^{-1},
\end{equation}
assuming a uniform field strength and density.  When perturbations of
wavelength $L$ are applied to a cold medium with the wavevector
perpendicular to the field, an analysis analogous to the classical
Jeans treatment (\cite{cha53}) shows that for $\beta \ll 1$ the 
perturbation is stable
when $\ta/\tc<1$ and unstable otherwise.  Thus for a
uniform field strength $B_0$, any cross-field column of $H$ less than
$ 4\times 10^{21} \cm^{-2} (B/10\mu \G)$ will be cross-field stable.
This borderline-stable column is consistent with the observed column
density seen for many clouds (\cite{lar81}) provided the field strength
$B\approx 25 \mu \G$.  The cross-field stability criterion $\ta/\tc <1$
is equivalent to the requirement that the mass-to-flux ratio in a
sphere carved out of this medium satisfy $M/\Phi<1/(3G^{1/2}) $.

For density perturbations with wavevectors parallel to the mean field
${\bf B}_0$, the Jeans gravitational stability analysis including just
thermal gas pressure is incomplete if, as has been proposed, \alf waves
provide the primary cloud support along the mean field.  In quasilinear
theory (cf. \cite{dew70}, \cite{mck95}), we can estimate the importance
of waves in opposing gravitational collapse along $\hat \bB_0$ by
performing a Jeans-type analysis with $\hat k\parallel \hat \bB_0$ and
pressure supplied by \alf waves according to $\Pw=P_{\rm wave, 0}
(\rho/\rho_0)^{1/2}$ (assuming $\ta \ll \tc,\ \ts$).  Here, $P_{\rm wave,
0}\equiv |\delta {\bf B}|^2/8\pi$ is the pressure in the fluctuating
field, and we assume \alf wavelengths short compared to the wavelength
of the density disturbance.  
Defining the Jeans number $\nj\equiv L/\Lj$, the criterion for
stability on a scale $L$ works out to be
\begin{equation}\label{PSJEANS}
\nj < \left(1 + {\Ew\over 4 \rb L \cs^2}\right)^{1/2}, 
\end{equation}
where the \alf wave surface energy density is $\Ew = \int\, dx{1\over
2}(\rho |{\bf \delta v}|^2 + |{\bf \delta B}|^2/4\pi) =
L\langle|{\bf \delta B}|^2\rangle/4\pi$.

An extension of the above quasilinear theory to the nonlinear regime
($E_W \gg \rb L \cs^2$) would argue that for arbitrary $\nj$, clouds
could be stabilized against collapse along $\hat \bB_0$ whenever there is
sufficient wave energy (in the absence of wave decay).  Properly, we do
not expect the analysis above to extend to the nonlinear regime
(indeed, while our nonlinear simulations do show correlations of wave
pressure and density, they do not obey any simple law like $\Pw\propto
\rho^{\gamma_p}$).  However, {\it any} argument that balances wave
energy in the slab against gravitational potential energy -- for
example, a virial analysis -- will yield the same scaling as equation
(\ref{PSJEANS}), since binding energy is proportional to $\nj^2\rb L
\cs^2$ (see eq. [\ref{EBIND}]).  The result of \cite{pud90} is
the same as equation (\ref{PSJEANS}) up to the factor $1/4$. 
Since our simulations begin from
uniform conditions, we find it convenient to use the quasilinear-theory
prediction of equation (\ref{PSJEANS}) as a reference point.  Of
course, this ``pseudo-Jeans'' analysis, or any analysis in terms of an
{\it initial} wave energy, becomes inapplicable if random motions
dissipate rapidly compared to the collapse timescale.

\subsection{Wave Dissipation Mechanisms}\label{S: dissipation}

The dominant linear damping mechanism in molecular clouds is ambipolar
diffusion (see \cite{mck93} and references therein).  Ambipolar
diffusion prevents propagation of \alf waves with frequencies $\omega_A =
{\bf k}\cdot {\bf \va}$ higher than a critical frequency $\omega_c =
2\nu_{ni}$, where the neutral-ion collision frequency is $\nu_{ni} \sim
n_i \, 1.5\times 10^{-9} \cm^3 s^{-1}$ (\cite{kul69}, \cite{nak84}).  
Estimates of
$\nu_{ni}$ indicate that waves can propagate at wavelengths well below
$0.01\pc$ (\cite{mye95}).  Ambipolar diffusion also damps \alf waves
with $\omega_A  < \omega_c$ at a rate $\dot \Ew/\Ew = -
\omega_A ^2/\nu_{ni}$.  
Writing the ionized fraction as $n_i/n_H=K n_H^{-1/2}$, the ambipolar-diffusion
damping time for wavelengths $\lambda$ is shorter than 
the gravitational collapse time $\tc$ when 
\begin{equation}
\lambda<14 \left({K\over 10^{-5}\cm^{-3/2}}\right)^{-1/2}
\left({\beta\over 0.01}\right)^{-1/2}\left({T\over 10\K}\right)^{1/2}
\left({n_{H_2}\over{10^2\cm^{-3}}} \right)^{-1/2}\pc.
\end{equation}
For ionization principally by cosmic rays at the fiducial
rate $10^{-17} s^{-1}$, $K\sim 10^{-5}$ (\cite{mck93}), 
which for typical parameters can imply significant frictional damping on 
dynamical timescales for all wavelengths.   Including the ionization produced 
by UV photons increases
$K$ and therefore decreases the frictional damping rate.
For $\beta<1$, fast MHD waves frictionally decay at a rate
comparable to \alf waves. Slow MHD (essentially acoustic) waves
suffer little frictional decay, but may be subject to other linear 
dissipation such as radiative damping.

Nonlinear ($\delta B/B \sim 1$) damping rates due to steepening of
compressive (fast magnetosonic) and transverse (Alfv\'enic) isolated
MHD wave trains have been compared by \cite{zwe83}.  The former
steepen at a rate $\lesssim k\delta v$, while the latter steepen at a
rate $\gtrsim k\delta v^2/v_A$; transverse waves damp more slowly
because pressure variations are second order in the wave amplitude
rather than first order.  Isolated wave trains deposit their energy in
shocks in approximately a wave-steepening time.  (Note that
parallel-propagating slow MHD waves with $\delta v/\cs \gg 1$ do not
exist; they shock immediately.)  Because \alf waves have the lowest
nonlinear damping rates, it has been suggested that the observed
supersonic linewidths owe their existence to these waves
(\cite{zwe83}).

Isolated circularly polarized \alf wave trains form a special case because
they are exact solutions to the compressible ideal MHD equations and 
therefore suffer no nonlinear steepening.  They are subject to a parametric
instability, however, that causes them to excite compressive and
magnetic disturbances over a range of $\omega$ and $k$ (\cite{gol78}).
The growth rate of the parametric instability is of order the wave
frequency when $\beta \lesssim 1$ and the dimensionless amplitude of
the wave is of order unity.  In the small-amplitude, low-$\beta$ limit
this instability becomes the well-known decay instability
(\cite{sag69}) in which a forward-propagating circularly-polarized \alf
wave decays into a forward-propagating acoustic wave and a
backward-propagating circularly-polarized \alf wave.  Circularly
polarized \alf waves cannot be relied upon, therefore, to provide
slowly damped motions in molecular clouds.

In truth, many waves are present in a molecular cloud simultaneously,
and interactions among waves are as important as the steepening
(essentially a self-interaction) of a single wave.   Wave interactions
transfer energy between scales; when energy is transferred to the
smallest scales, dissipation occurs.  The energy spectrum that develops
depends of course on the allowed wave families in the fluid.  The most
familiar example of this process is the Kolmogorov cascade in an
incompressible, unmagnetized fluid, which leads to an energy spectrum
$dE(k)/dk\propto k^{-5/3}$.  For magnetized, incompressible fluids
($\cs/\va\gtrsim 1$), a theory of weak turbulence based on coupling of
shear \alf waves has recently been developed by \cite{sri95} (SG).  The
cascade to small scales occurs predominantly perpendicular to the mean
magnetic field, but the weak theory becomes invalid at large $k_\perp$
due to increasing nonlinearity.  A theory of critically-balanced strong
turbulence, again for incompressible media, has been proposed by
\cite{gol95} (GS).  

Turbulence in molecular clouds lies in a rather different part of
parameter space than that considered by Goldreich \& Sridhar, so their
theory cannot be directly applied.  The SG+GS theory was developed for
application to scintillation in the ionized interstellar medium, where
$c_s/v_A \gtrsim 1$.  In molecular clouds, by contrast, self-gravity
and efficient cooling have conspired to produce a highly compressible
medium ($\cs/\va \ll 1$) in which turbulence is probably strong
($\delta v/\va\sim 1$) at the largest scales. In low-$\beta$ media,
compressive disturbances are easily excited and damp rapidly via shocks.
In general, a low-$\beta$ plasma can be expected to generate both an
anisotropic shear-\alf cascade, similar to that studied by SG+GS, and
compressive disturbances and shocks (cf.  Ghosh \& Goldstein (1994)).
Only with high-resolution two- or three- dimensional numerical studies
of compressible MHD turbulence will it be possible to delineate the
circumstances when one or the other mechanism provides the primary
dissipation for Alfv\'enic disturbances in conditions appropriate for
molecular clouds.  In this first study, we have adopted a compressible
equation of state but use the the simplest possible geometry (slab
symmetry) that allows for transfer of \alf-wave energy to compressive
motions.  Insofar as our model allows for compressibility but not a
perpendicular \alf-wave cascade, our study is complementary to that of
Goldreich \& Sridhar.

\section{Basic Equations, Numerical Method, and Tests}\label{NUMER}

In this paper we consider the simplest possible system in which nonlinear
MHD waves and gravity can interact.  We solve the equations of
self-gravitating, compressible, ideal MHD:
\begin{equation}\label{BAS1}
{\del\bv\over{\del t}} + (\bv\cdot\nabla)\bv =
	-{\nabla P\over{\rho}} - {\nabla \bB^2\over{8\pi\rho}}
	+ {(\bB\cdot\nabla)\bB\over{4\pi\rho}} - \nabla\phi,
\end{equation}
\begin{equation}\label{BAS2}
{\del\rho\over{\del t}} = -\nabla\cdot(\rho\bv),
\end{equation}
\begin{equation}\label{BAS3}
{\del \bB\over{\del t}} = \nabla \times (\bv \times \bB),
\end{equation}
\begin{equation}\label{BAS4}
\nabla^2 \phi = 4 \pi G\rho.
\end{equation}
The equation of state is isothermal,
\begin{equation}
P = c_s^2 \rho,
\end{equation}
with $c_s = 1$ throughout.  

We assume ``slab symmetry'', that is, all variables are a function of
one independent spatial variable $x$ and time $t$.  Thus derivatives in
the transverse ($y$ and $z$) directions are zero, and the only
derivatives that survive in equations (\ref{BAS1}) to (\ref{BAS4}) are
in the longitudinal ($x$) direction.  Transverse velocities $\bf
v_\perp$ and magnetic fields $\bB_\perp$ also survive and vary with
$x$, making our scheme ``$1 + 2/3$ D''.  Notice that because of the
symmetry and the constraint $\nabla\cdot\bB = 0$ we have $B_x(x,t) =
const.$ The boundary conditions are periodic in a slab of thickness $L
= 1$, i.e. a fluid element leaving the model at $x = -L/2$ reenters at
$x = L/2$.  For periodic boundary conditions, $\rho\rightarrow \rho -\bar\rho$
on the right-hand-side of equation (\ref{BAS4}) (see \S \ref{GRAVDECAY}).

Our numerical method is a one dimensional implementation of the ZEUS
code described by \cite{sto92a} and \cite{sto92b}.  The hydrodynamical
portion of the code is a time-explicit, operator-split finite
difference algorithm on a staggered mesh.  Density and internal energy
are zone-centered, while velocity components dwell on zone faces.  The
magnetic portion of the code uses the Method of Characteristics to
evolve the transverse components of the magnetic fields and velocities
in a manner that assures the successful propagation of \alf waves.

The gravitational acceleration, when used in our model, is 
calculated by taking the Fourier transform of the density and, for each
component of the potential $\phi_k$, setting $\phi_k = 4\pi G\rho_k
(\Delta x)^2/(2 \cos(k\Delta x) - 2)$, where $\Delta x$ is the zone
spacing.  This gravitational kernel ensures that Poisson's equation is
satisfied in finite-difference form.  The gravitational acceleration is 
calculated by taking the inverse transform and setting $g_i =
-(\phi_{i+1/2} - \phi_{i-1/2})/\Delta x$.  We have confirmed that 
selected spatial modes obey the Jeans dispersion relation as a test 
of the gravitational portion of the code.  

As a further comparison with linear theory, we have verified the
scalings $\Pw\propto\rho^{3/2}$ and $\Pw\propto \rho^{1/2}$ for the
\alf wave pressure in, respectively, an adiabatically contracting cloud
and a wavetrain propagating along a density gradient (\cite{mck95}).
To verify the first scaling, we imposed a slow contraction of the
coordinate system in a simulation containing a low-amplitude \alf wave
and found good agreement with the adiabatic wave amplitude-density
relation.  To verify the second scaling, we imposed an external
gravitational potential to set up a density gradient and forced a
low-amplitude \alf wave at the bottom of the potential well.  Again, we
found that the wave amplitude for the outward-propagating wavetrain
agreed well with the linear ``polytropic'' theory.

While our numerical method thus produces good agreement with linear 
theory, the experiments we present below involve the evolution
of highly nonlinear MHD systems.  It is hard to find good code tests
for such systems, since there are few exact solutions known.  Tests
that exercise an MHD code's ability to handle the full family of MHD
discontinuities have been described by \cite{ryu95}; we have performed
these tests and obtained satisfactory results.

Another direct and germane nonlinear test of the code is provided by
the parametric instability of circularly polarized \alf waves in a
compressible fluid described by \cite{gol78}.
We have simulated the instability of circularly polarized
\alf waves and verified that the growing disturbances obey Goldstein's
analytic dispersion relation.  
We have also simulated the steepening of an elliptically polarized \alf
wave and found good agreement with the calculations of \cite{coh75}.

\begin{figure}
\plotone{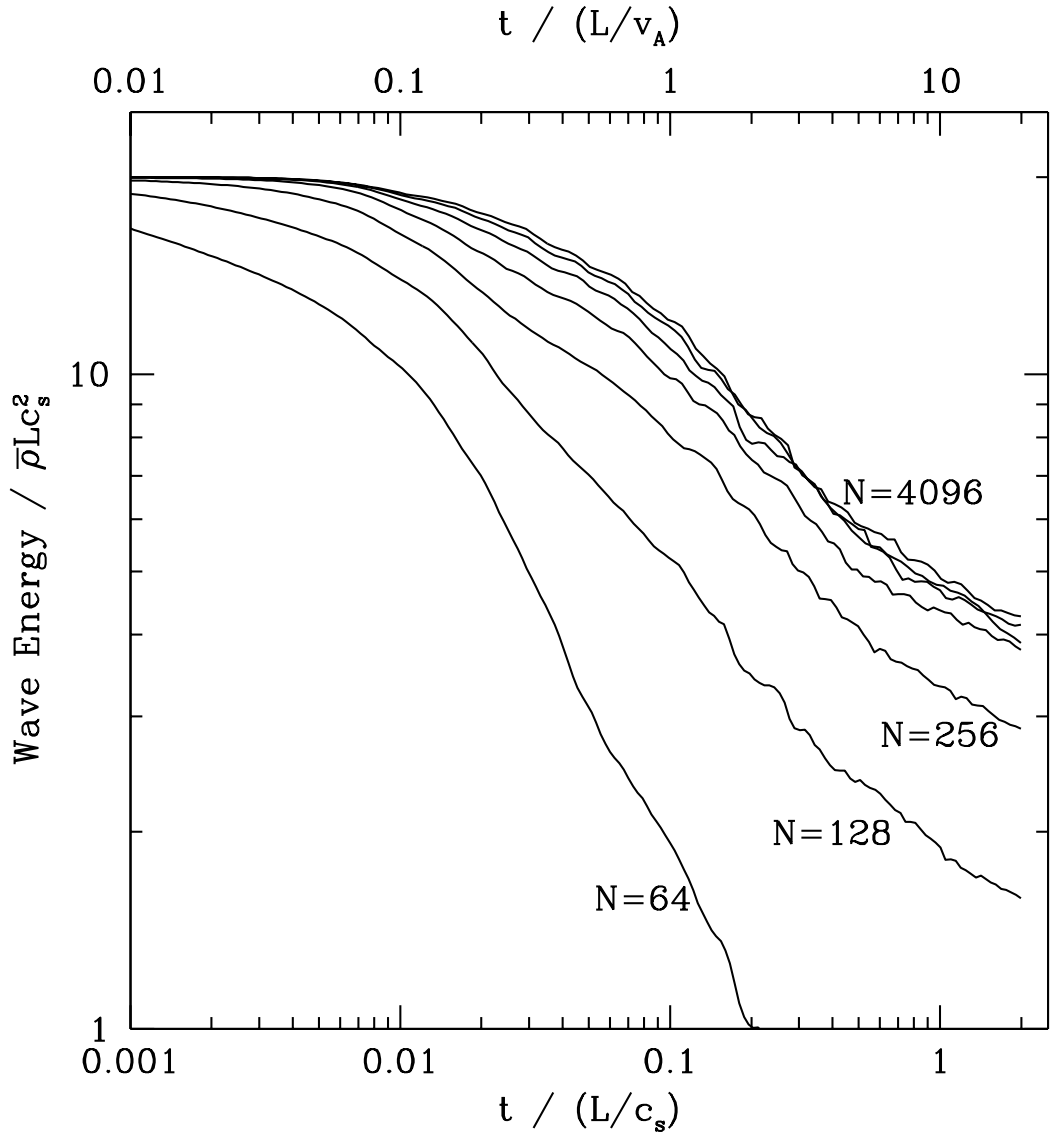}
\caption{
A convergence test.  The figure shows the evolution of the
sum of the magnetic and kinetic energy in simulations that begin
with an initially random transverse field and velocity, but with
different numerical resolution.  This shows that the evolution
does not depend on numerical effects for large enough resolution.
}
\end{figure}

Finally, we have convergence tested the code for many of the
simulations described in this paper.  An example is shown in Figure 1,
which displays the evolution of ``wave'' energy (kinetic and magnetic
energy) in a suite of simulations that begin with a ``random'' 
set of transverse velocities and magnetic fields (see \S 4.1 for details 
of the initial conditions).  The simulations vary
only in their numerical resolution; the initial conditions are
identical in the sense that their Fourier transforms are the same.
Evidently as resolution increases the energy evolution converges.  This
is because numerical diffusion decreases as resolution increases, while
the real sources of dissipation -- isothermal shocks-- become increasingly well
resolved.  Provided we use sufficient resolution (of order $512$
zones in this case), the energy evolution will not depend qualitatively
or quantitatively on numerical effects.

\section{Simulations}\label{S: Simulations}

We have performed several different types of simulations using the
basic model described in the last section: a periodic, one-dimensional
system with magnetic fields and an isothermal equation of state.  This
model may be thought of as mocking up a piece of a molecular cloud
that is dominated by a mean field lying in the longitudinal ($x$)
direction, although it also serves as perhaps the simplest possible
context in which to study the interaction of \alf waves and
self-gravity.  All the simulations begin with a uniform initial
density $\rho = 1$, zero longitudinal velocity ($v_x = 0$), and zero
mean transverse field ($\bar{B}_y = \bar{B}_z = 0$; we have verified
that small mean transverse fields do not change our results
qualitatively).  Each simulation can be characterized by the
(unchanging) strength of the mean longitudinal magnetic field $B_x$,
expressed in terms of the parameter $\beta \equiv \cs^2/\va^2 \equiv
\cs^2/(B_x^2/4\pi \rho)$ (see eq. [\ref{betadef}]; we include only the
mean field component $B_x$ in $\va$).  The numerical resolution for
the ``standard'' runs described below is $2048$ zones; all other runs
were done with $512$ zones.  Four different types of simulations are
discussed in the subsections below.  They are: the free decay of an
initial wave spectrum (\S \ref{FREEDECAY}); a forced equilibrium in
which the model is stirred at the same rate that it dissipates energy
(\S \ref{SUBSFEQR}); the free decay of an initial wave spectrum in the
presence of self-gravity (\S \ref{GRAVDECAY}); and a forced
equilibrium in the presence of self-gravity (\S \ref{GRAVFORCED}).

A useful diagnostic for the simulations described below is what
we shall call the {\it wave energy}.  It is defined as the sum
of the kinetic energy and the energy of the perturbed magnetic field, 
$\Ew=\Ek+\Eb$, where
\begin{equation}
\Ek\equiv \int_{-L/2}^{L/2} 
dx {1\over 2}\rho (|{\bf v}_\perp|^2 +v_x^2)
\end{equation}
and
\begin{equation}
\Eb\equiv \int_{-L/2}^{L/2} 
dx {|{\bf B}_\perp|^2\over 8\pi}.
\end{equation}
Notice that $\Ek,\Eb,$ and $\Ew$ are in units of energy per unit surface 
area because of the slab geometry.
For reference, the rms velocity and magnetic field perturbations satisfy
$\<|\delta \bv/\cs|^2\>^{1/2} = (2\Ek/\rb L\cs^2)^{1/2}$ and 
$\<|\delta \bB/B_x|^2\>^{1/2} =\beta^{1/2}(2\Eb/\rb L\cs^2)^{1/2}$.  The 
definition of $\beta$ implies that $v/\va=\beta^{1/2} v/\cs$ for any $v$.  
Time is given in units of the sound-crossing-time $\ts\equiv L/\cs$, and 
the \alf-wave crossing time $\ta=L/\va=\ts\beta^{1/2}$. For self-gravitating 
simulations, the collapse time is $\tc=\ts/\nj$, where
$n_J \equiv L/L_J$  (see eqs. [\ref{tsound}], 
[\ref{tcollapse}], and [\ref{talfven}]).

\subsection{Free Decay}\label{FREEDECAY}

First consider a non-self-gravitating model containing a spectrum of
waves in the initial conditions that are allowed to freely decay
thereafter.  The ``standard'' run has $\beta=0.01$ and initial wave energy
$\Ew=2\Ek=2\Eb=\ 100 \rb L c_s^2$.  With this initial energy, 
the initial field line distortions have a dimensionless amplitude
$\<|\delta \bB/B_x|^2\>^{1/2} =1$.

The initial transverse velocities and magnetic fields are a ``random''
superposition of parallel-propagating \alf waves.  They are drawn from
a Gaussian random field with power spectrum $\langle|\bv_{\perp,k}|^2\rangle =
\langle|\bB_{\perp,k}|^2/{4\pi\rb} \rangle \propto k^{-2}$ for 
$2\pi/L < |k| < 32 (2\pi/L)$.  This power spectrum is special in that it 
approximates the natural
power spectrum to which other initial power spectra decay, according to
our numerical experiments, and also in that it corresponds to a
velocity-dispersion/size relation $\sigma_v(R)^{1/2}\propto
R^{1/2}$ in one dimension which is consistent with one of Larson's laws
(cf. \cite{lar81}; \S \ref{S: cloud properties}).  The power spectrum
is steeply declining, so almost all the energy is at the largest
scales.

\begin{figure}
\plotone{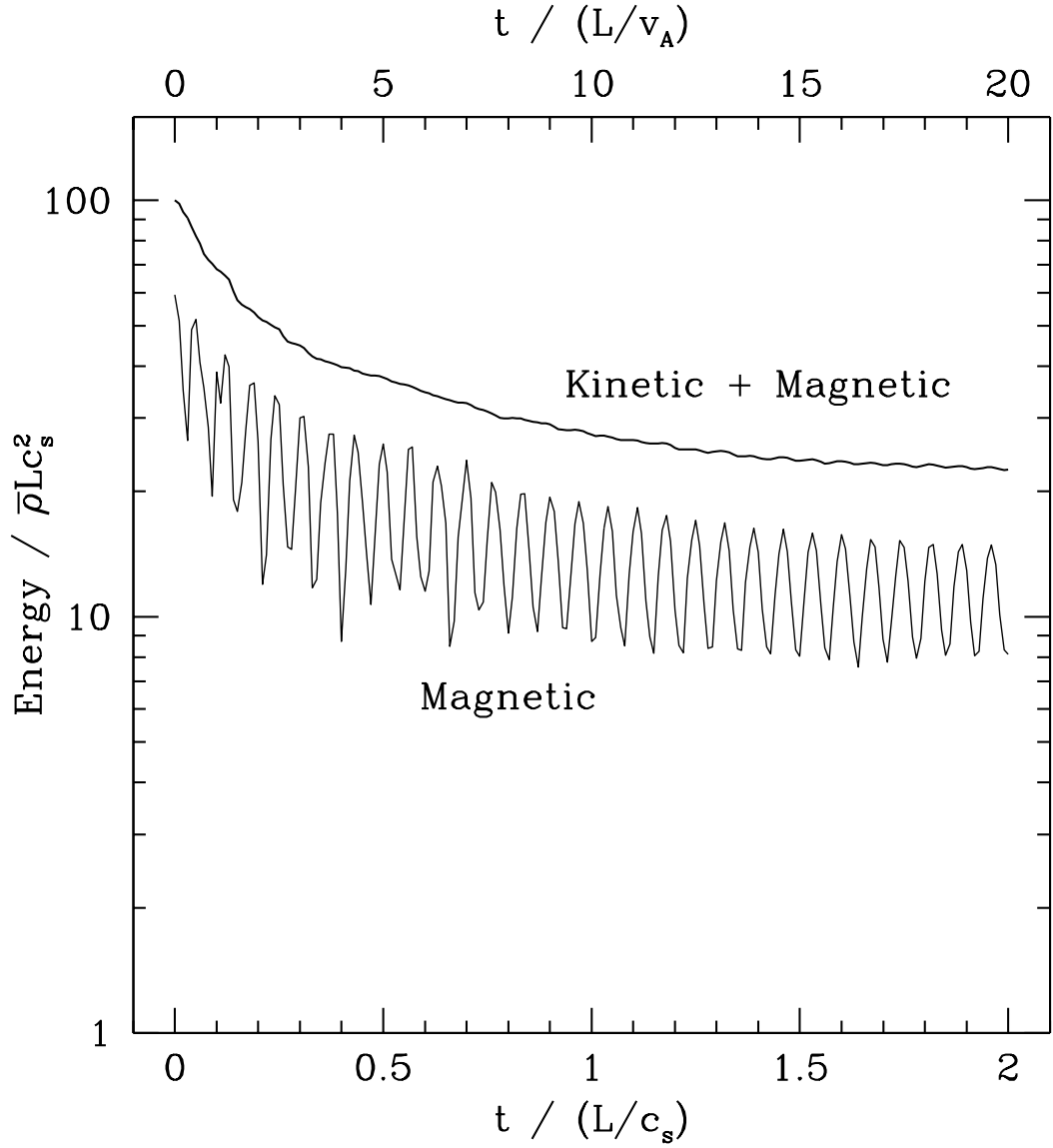}
\caption{
Time evolution of the transverse magnetic + kinetic energy and
the magnetic energy in the ``standard'' decay simulation.
}
\end{figure}

The evolution of the total wave energy and the magnetic energy are
shown in Figure 2.  The initial transverse magnetic fields give rise to
a nonuniform magnetic pressure that is large compared to the gas
pressure.  This in turn produces longitudinal accelerations, in
addition to the transverse accelerations caused by magnetic tension.
The transverse kinetic and magnetic energies remain nearly in
equipartition and oscillate out of phase, yielding a smoothly-evolving
total wave energy;  the longitudinal kinetic energy is only a few
percent of the total.  The wave energy declines to half its initial
value by $t = 0.2 \ts=2.0 \ta$.  The transverse motions and fields lose
their energy when the magnetic pressure due to the transverse fields
does $P dV$ work on the fluid,
\begin{equation}
{d\over{dt}}\int_{-L/2}^{L/2} 
\,dx\left({1\over{2}}\rho |\bv_{\perp}|^2
	+ {|\bB_\perp|^2\over{8\pi}}\right) =
\int_{-L/2}^{L/2} \,dx\,v_x {\del\over{\del x}}
	\left({|\bB_\perp|^2\over{8\pi}}\right),
\end{equation}
producing longitudinal motions that dissipate in shocks.  Toward the
end of the simulation almost all the wave energy is concentrated at the
largest scales, thus giving the regular oscillations in transverse
magnetic energy seen in Figure 2.  

\begin{figure}
\plotone{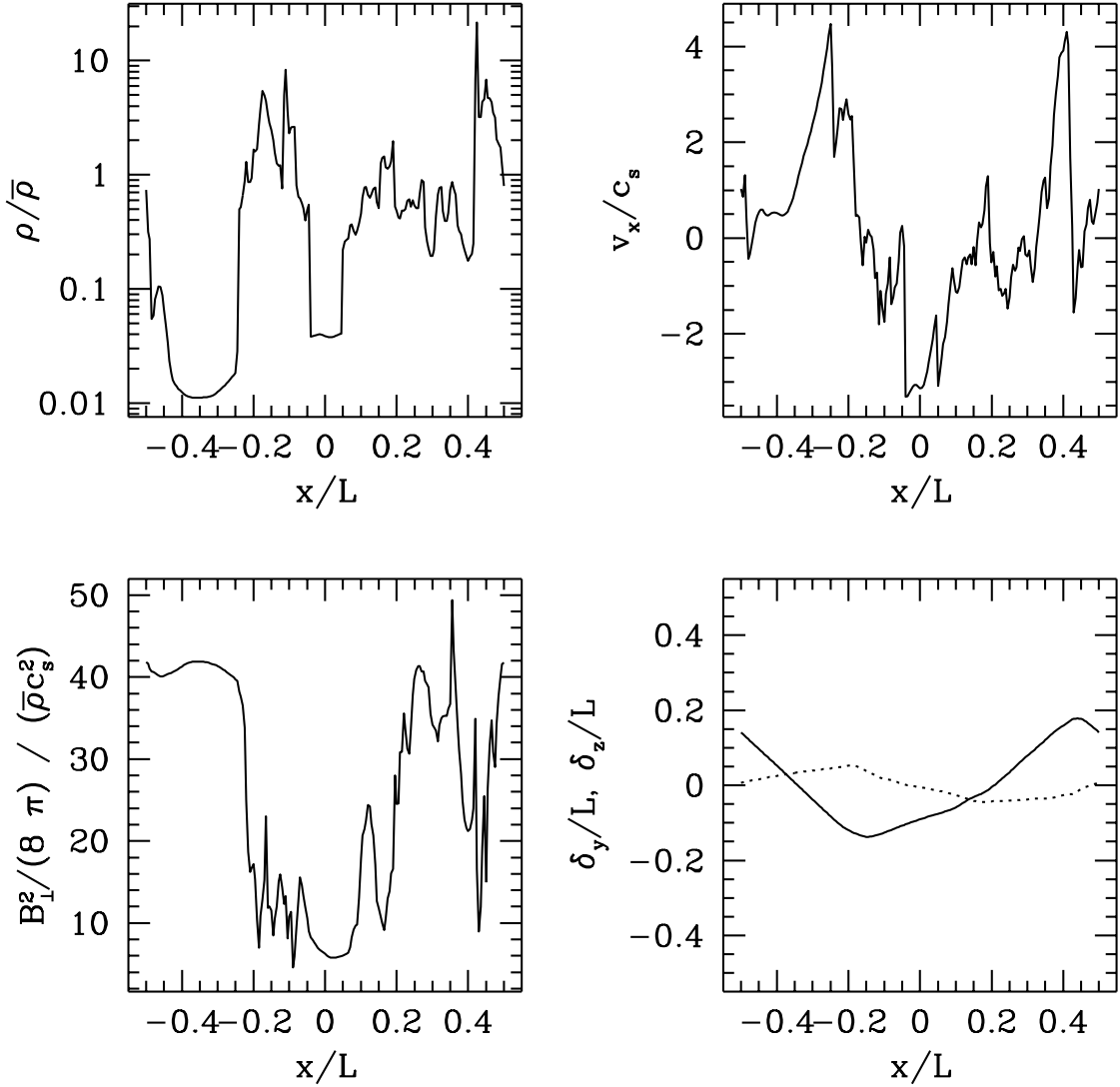}
\caption{
Portrait of density $\rho$, longitudinal velocity $v_x$, transverse
magnetic pressure $B_\perp^2/8\pi$, and transverse field line
displacements $\delta_y$, $\delta_z$ in the ``standard'' decay
simulation at time $t=0.5\ts=5\ta$.
}
\end{figure}

Figure 3 shows density, longitudinal velocity $v_x$, magnetic pressure,
and the field line geometry of the simulation at $t = 0.5 \ts=5\ta$.
There are large density contrasts present, with densities ranging over
three orders of magnitude.  Numerous shocks provide the dissipation.
Notice that the magnetic pressure dominates the gas pressure (the gas
pressure is numerically equal to the density, since $c_s^2 = 1$).  The
field line structure in the $x-i$ planes ($i = y,z$) is shown in the
lower right panel of the figure, with the $i$ field displacement
$\delta_i$ defined by
\begin{equation}
\delta_i(x) = \int^x\, dx\,{B_i\over{B_x}} - const.,
\end{equation}
where the constant is set so that $\bar{\delta_i} = 0$.  Because most
of the energy remains in the largest-scale Fourier components of the
field, the field displacements are well ordered, with one maximum and
one minimum.  The field lines are nearly straight, and hence
force-free, in between the kinks at density maxima.

\begin{figure}
\plotone{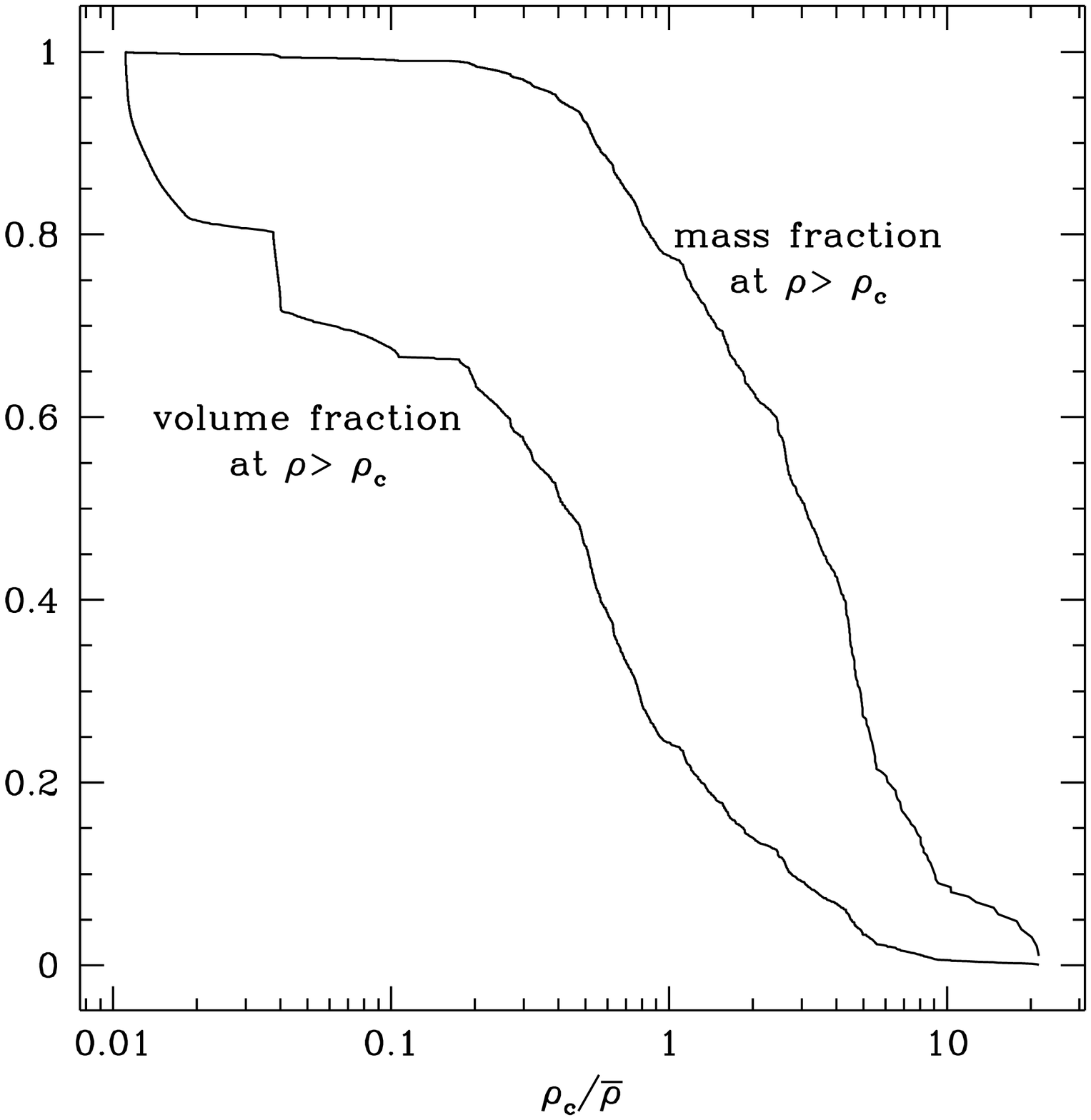}
\caption{
Volume and mass fractions above density enhancement
level $\rho_c/\rb$, for the ``standard'' decay simulation at time 
$t=0.5\ts=5\ta$.
}
\end{figure}

The variations in density at $t = 0.5\ts=5\ta$ can be further
characterized by computing the fractional volume and fractional mass at
$\rho > \rho_c$ (Figure 4).  The figure shows that $10\%$ of the
mass (volume) is at $\rho > 9.1\rb$ ($2.7\rb$), $50\%$ of the
mass (volume) is at $\rho > 3.1\rb$ ($0.42\rb$), and $90\%$ of the
mass (volume) is at $\rho > 0.55\rb$ ($0.012\rb$).  Thus the mass is
concentrated in dense regions, while a significant fraction of the
simulation volume is nearly empty.  These results are typical of all 
our simulations, although density contrast increases significantly in 
the self-gravitating simulations discussed below.  

At time $t = 0.5\ts=5\ta$, the best-fit slope $s$ to the combined power
spectrum $|\bv_{\perp,k}|^2 + |\bB_{\perp,k}|^2/{4\pi\rb}\propto
k^{-s}$ yields $s=2.3$ between $k/(2\pi/L)= 1$ and $100$.  The slope
varies rapidly, however, with $\<s\> = 2.4 \pm 0.16$, where $\<\>$
indicates an average over time.  By smoothing the $\bv_\perp$ data over
Gaussian windows of varying width and averaging over the box, and over
time, we can compute a simulated ``linewidth-size'' relation (see \S
4.2).  Averaged over time, the best fit relation is $\sigma_v(R)\propto
R^{0.7}$.

A vital question for the structure of molecular clouds is whether the
wave energy concentrates in the regions of highest density.  This can
be answered by comparing $|\delta \bB|^2$ with $\rho$ after
smoothing both on a scale $\lambda$.  In our
standard decay simulation, magnetic pressure and density are weakly
{\it anti}correlated in a time-averaged sense, with slope $|\delta \bB|^2
\propto \rho^{-0.1}$ for essentially all smoothing scales $\lambda$ (in
\S \ref{GRAVDECAY} we show that self-gravity changes this
correlation).  Thus magnetic pressure plays a role in confining the
``clumps''.  But the clumps are not in equilibrium: they form,
disperse, and accelerate, and material migrates from one clump to
another.  We do find that clump masses grow over time, however, as
energy is transferred to larger scales and more coherent motions.

\begin{figure}
\plottwo{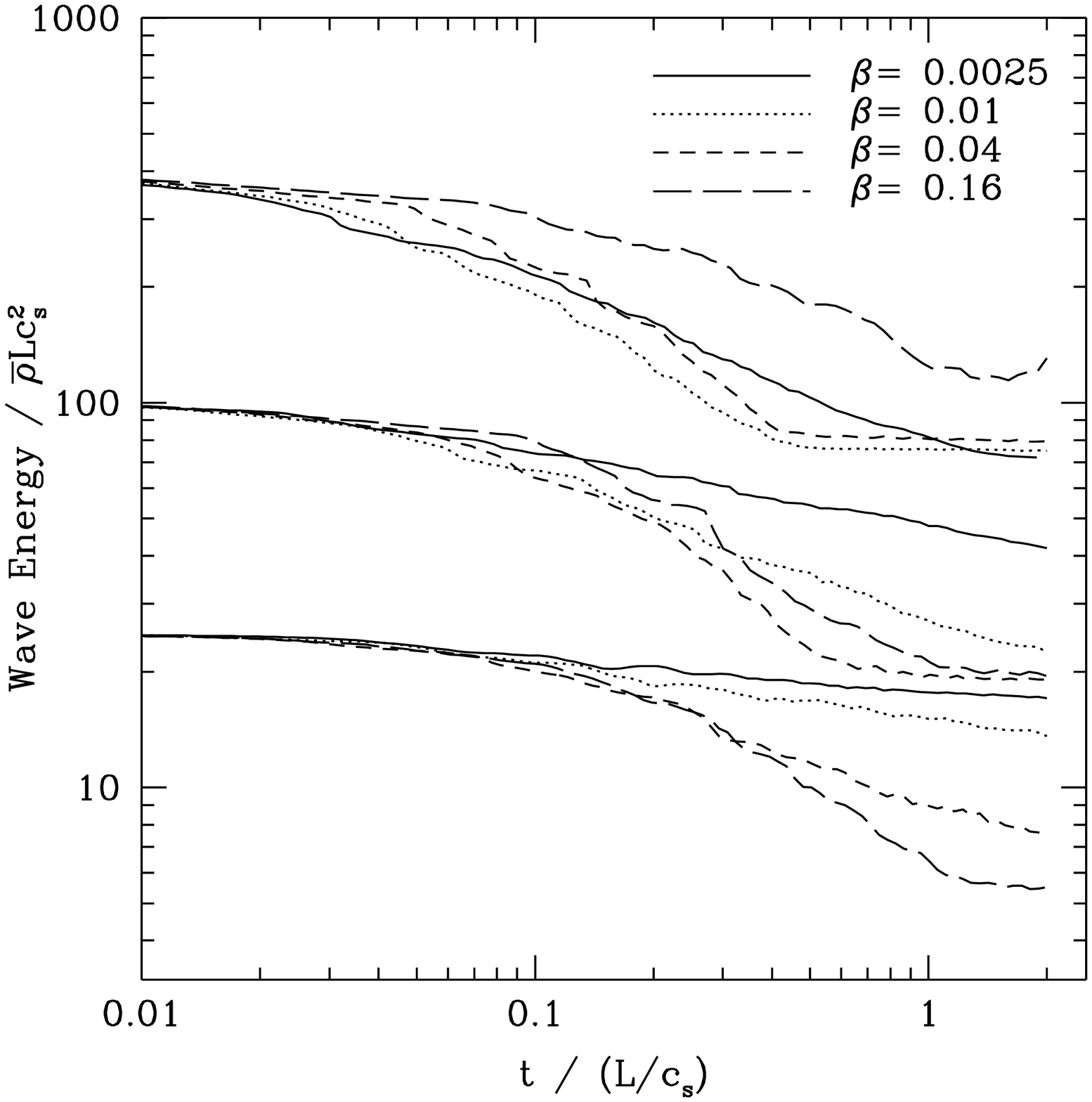}{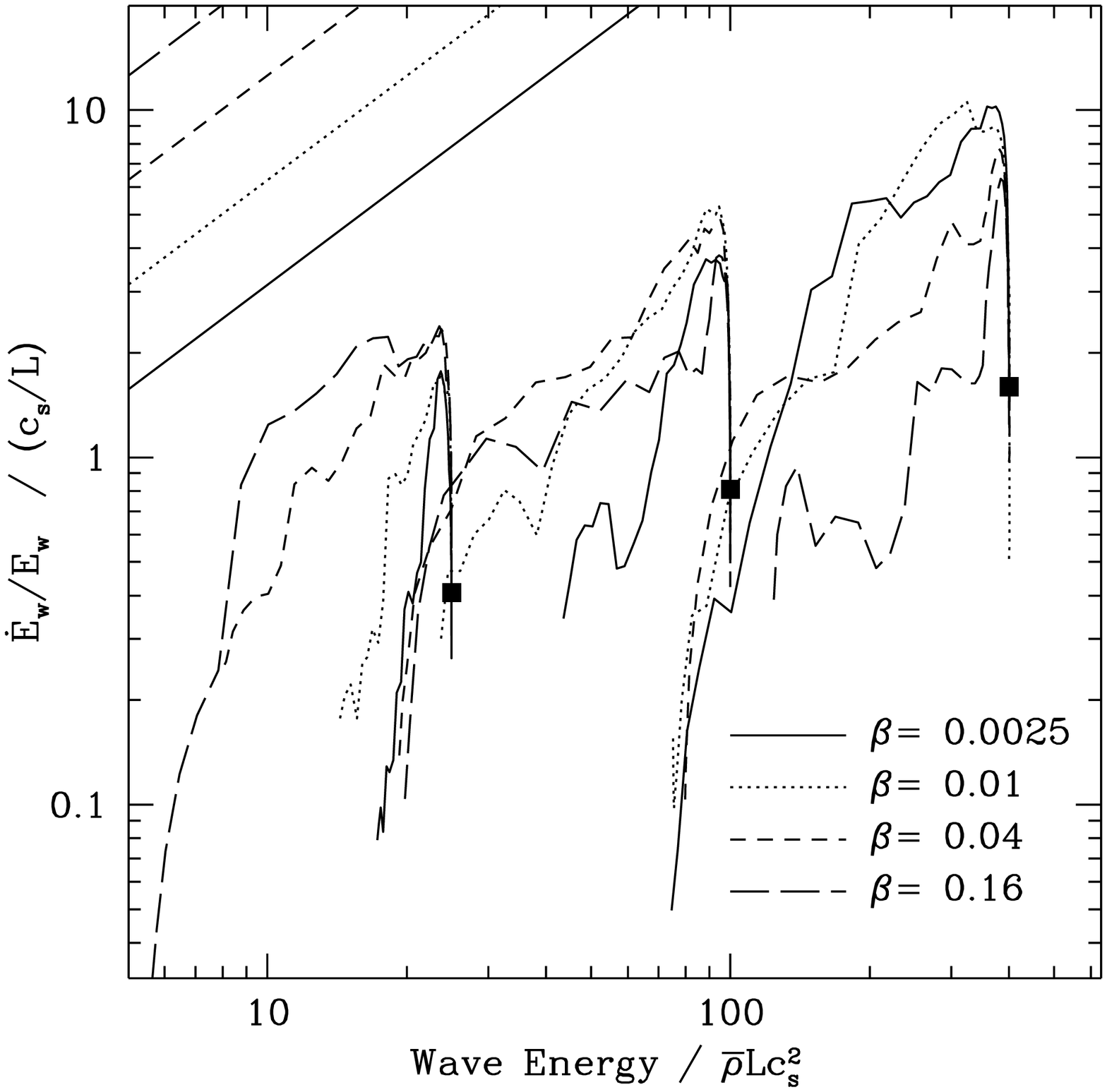}
\caption{
(a) Time evolution of the magnetic+kinetic energy in a series of
decay simulations with initial wave energy (transverse kinetic $+$ 
transverse magnetic) of $25, 100,$ and $400$, and initial
$\beta \equiv \cs^2/\va^2 = 0.04, 0.01,$ and $0.0025$.
The \alf time is $\ta/\ts=0.05, 0.1, 0.2, 0.4$ for
$\beta=0.0025, 0.01, 0.04, 0.16$, respectively.
(b) The dissipation rate $-\dot{\Ew}/\Ew$ 
as a function of $\Ew$ in the same simulations as (a).  The 
simulations begin in the neighborhood of the solid black square.
The dissipation rate initially increases, then decreases, as the 
simulations evolve and energy decays.
Toward the end of many of these simulations the dissipation rate is
sharply reduced because most of the power is in the lowest order
($k = 2\pi/L$) Fourier mode, which decays very slowly.
The scale on the ordinate gives the decay rate in units $\ts^{-1}$.  For
$\beta=0.0025, 0.01, 0.04, 0.16$, a decay rate 
$\ta^{-1} = 20, 10, 5, 2.5 \ts^{-1}$, respectively.  The straight lines
show single wave steepening rates based on Cohen \& Kulsrud (1974).
}
\end{figure}

We have surveyed free decay models with the same initial power spectrum
over a variety of $\beta$ and initial $E_W$.  The evolution of the wave
energy in a selected sample of these simulations is shown in Figure
5a.  The initial
dimensionless wave amplitudes range from $\<|\delta \bB/B_x|^2\>^{1/2}
= 1/4$ to $8$.  Considering the subset of simulations where the initial
rms field perturbations $(\beta\Ew/\rb L\cs^2)^{1/2}$ are $2$ or less,
for a given initial $\Ew$ the decay is slower as the field strength
increases ($\beta$ decreases).  For this same set of simulations, when
we consider decays at a given $\beta$, the fractional energy loss at
$\ta$ increases as initial $\Ew$ increases.

Some of the simulations with initial wave energy of $400\rb L c_s^2$
stop decaying near the end of their run.  This is because they are
able to make their way into a special dynamical state.  In this
special low-decay state, all the matter is concentrated into two
narrow sheets that oscillate transversely.  This special state is
artificial, because its stability depends on the boundary conditions.
We have tried duplicating the special state into a region of length $2
L$, so that there are four narrow sheets, and introducing some small
perturbations.  We find that the system becomes unstable and decays to
a new special state with only two sheets.  The special state may also
suffer higher dimensional instabilities which cannot be represented in
the present simulations.  While our simulations display a common trend
toward concentration of energy and structure at the largest available
scales, it seems unlikely that suitable conditions for the persistence
of this state will be found in nature.

It is not possible to define a ``decay rate'' from the
free decay simulations because the decay rate depends both on the
energy and on the internal state of the system.  This is made clear
in Figure 5b, which shows the instantaneous decay rate $-\dot{\Ew}/\Ew$
for the same runs shown in Figure 5a.  The heavy black squares indicate
where the simulations begin; the decay rates quickly rise, then the
energies and the decay rates fall.  Except for initial 
transients in a few of the high-$\beta$, high-$\Ew$ cases, decay rates lie
below $\ta^{-1}$.

In Figure 5b, the straight lines show the nonlinear decay rate $-\dot
\Ew/\Ew = 2\pi (\Ew/\rb L \cs^2) \beta^{1/2} \ts^{-1}$ based on the
single-wave steepening calculation of \cite{coh75}, which has sometimes
been used to estimate the dissipation rate within molecular clouds.
This nonlinear steepening rate overestimates the decay rates we find by
an order of magnitude or more, and does not display the same scaling as
either the peak decay rates for different runs at a given $\beta$, or
the evolutionary decay rate for any given run.  Indeed, one would not 
expect the Cohen \& Kulsrud  estimates to agree with our
simulations, since: (a) the simulations contain a spectrum of waves,
while the CK estimate is for a monochromatic wave train; (b) the disturbances
in our simulations are highly nonlinear at the outset--
they do not gradually steepen from a near-mode of the plasma.  To the
extent that a single decay rate $\gamma$ can characterize these
freely-decaying systems as a function of $\Ew/\rb L\cs^2$ and $\beta$,
we note that the upper envelope of the decay rate in Figure 5b has
better consistency with the scalings and approximate magnitude of the
decay rate measured in the steadily forced simulations of the next
subsection.

\begin{figure}
\plotone{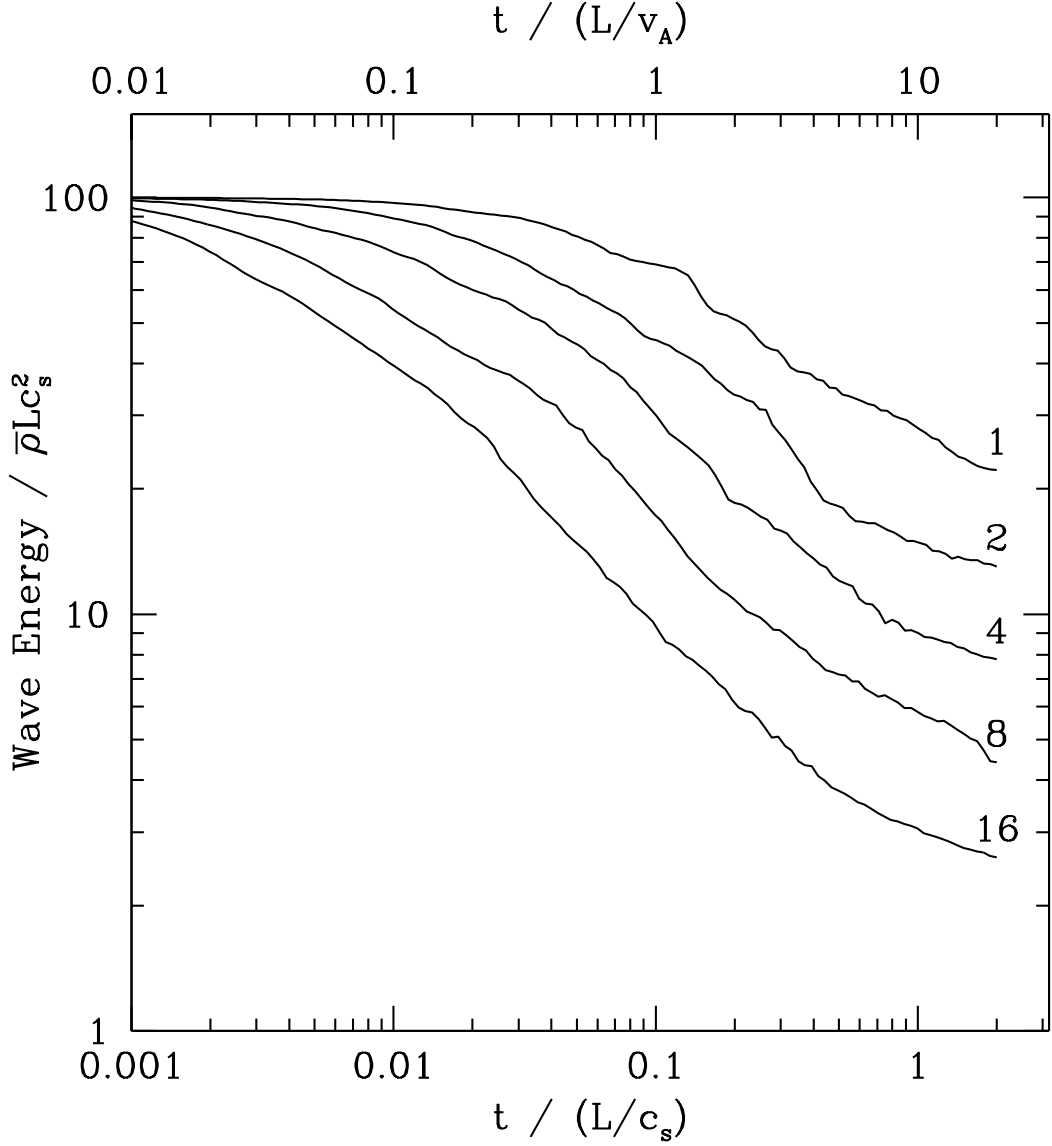}
\caption{
Time evolution of the wave energy (kinetic + transverse magnetic) in
decay simulations where the initial energy power spectra extend between
$k_{\rm min}/(2\pi/L)=1, 2, 4, 8, 16$ (see labels) and $k_{\rm
max}/(2\pi/L)=64$.  The figure demonstrates that decay proceeds more
quickly when wave energy is stored in smaller-scale, higher-frequency
oscillations.
}
\end{figure} 
 
The dissipation rate depends strongly on the initial power spectrum.
We have performed a series of simulations with 
$|\bv_{\perp,k}|^2 =
|\bB_{\perp,k}|^2/(4\pi\rb) \propto k^{-2}$ for 
$k_{\rm min} < k <k_{\rm max}$ and $0$ elsewhere, 
where $k_{\rm min}/(2\pi/L) = 1,2,4,8,$ and $16$ and 
$k_{\rm max}/(2\pi/L) = 64$.
In all cases $\beta=0.01$.  The time evolution of the wave energy in
these simulations is shown in Figure 6; evidently the smaller the
scale at which the energy is injected the more rapid the
dissipation.  The peak decay rates are $\simeq 0.2 \omega_A$.  Our
examination of the evolution of the density and field structure, and
the power spectra of the transverse velocities and field (not
shown), shows that while energy is lost overall, some fraction of
the energy is transferred to modes with $k$ smaller than the initial
$k_{\rm min}$.  The final power spectra have $|\bv_{\perp,\,k}|^2,
|\bB_{\perp,\,k}|^2\propto k^{-2.5\pm 0.2}$  
independent of where the cutoff
$k_{\rm min}/(2\pi/L)$ was imposed in the input spectrum.  
We have also initiated decay simulations with alternative spectral slopes, and
found that they too evolve toward spectra with slopes in the range $-2$ 
to $-2.5$.

\subsection{Forced Equilibrium Runs}\label{SUBSFEQR}

The MHD waves in molecular clouds may be trapped and preserved from the
time of cloud formation, and the free decay simulations are designed to
model that possibility.  An alternative, however, is that they are
continuously driven by processes inside the cloud, such as winds from
young stars (\cite{nor80}).  In this section we consider stochastic
driving of transverse motions as a model for this process.

The forcing is accomplished by introducing a pattern of transverse
velocity perturbations $\delta\bv_\perp$ at regular time intervals
$\delta t$.  The velocity perturbations have the following properties: (1)
$|\delta\bv_{\perp,\,k}|^2 \propto k^4 \exp(-4 k/k_{\rm pk})$, so that
the power spectrum is peaked at $k = k_{\rm pk}$; (2) the
phases of the Fourier components of the perturbation are random; (3)
the power spectrum is normalized and the mean component of the
transverse velocities set so that
\begin{equation}
\delta E = \int\,dx {1\over{2}}\rho \left(|\bv + \delta\bv|^2 
	- |\bv|^2\right) = const.
\end{equation}
and 
\begin{equation}
\int\,dx\rho\delta\bv = 0.
\end{equation}
The forcing interval $\delta t$ is typically set to be one one-hundredth 
of a sound crossing time, while $k_{\rm pk}$ is $8 (2\pi/L)$ unless stated
otherwise.

\begin{figure}
\plotone{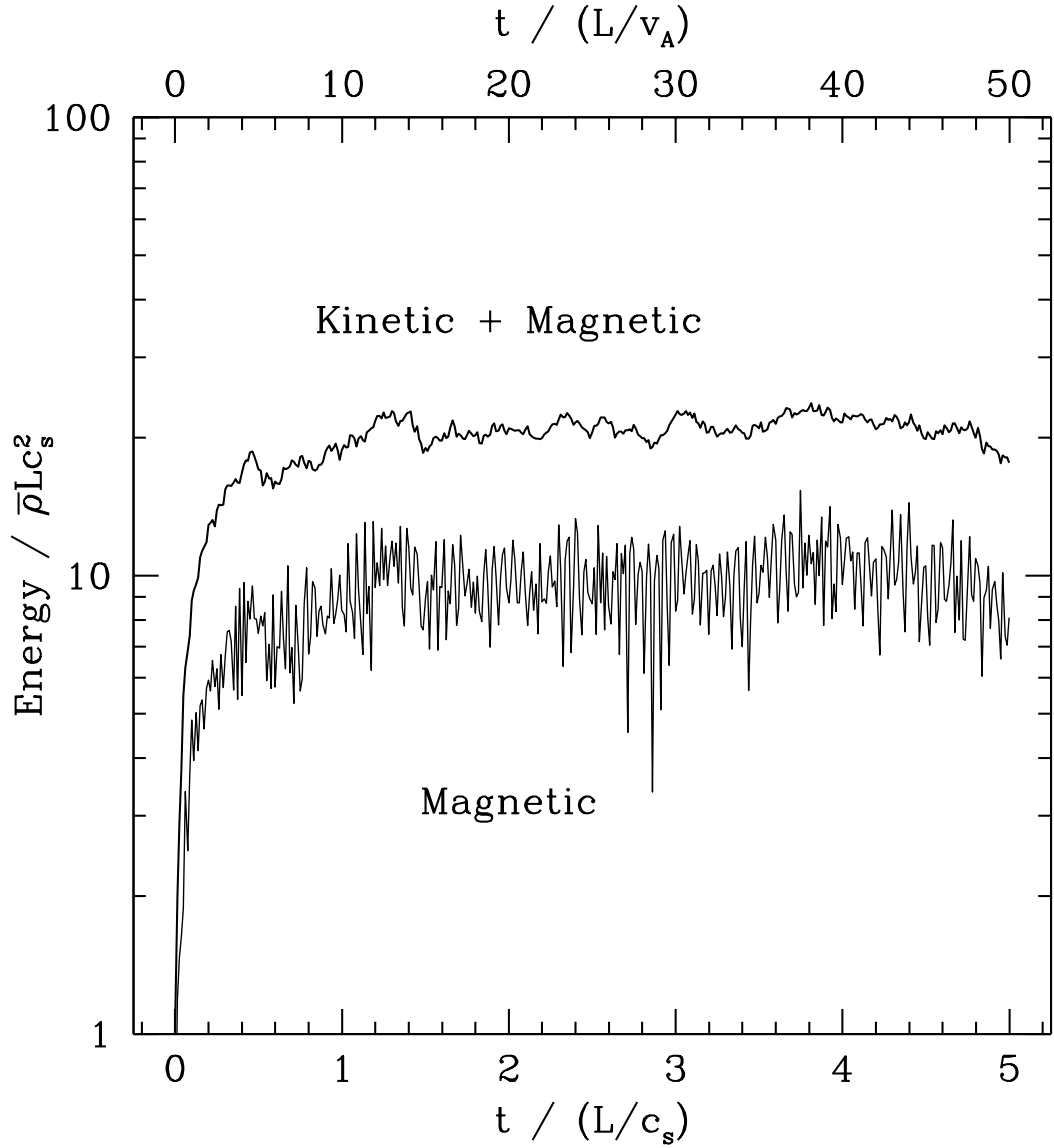}
\caption{
Time evolution of the magnetic+kinetic energy and the magnetic energy
in the ``standard'' forced run.  The forcing is accomplished by adding
random transverse velocities at fixed time intervals.
}
\end{figure}

The evolution of the standard forced run in shown in Figure 7, for
which $\beta = 0.01$ and the forcing power $\delta E/\delta t = 100 \rb
c_s^3$.  The model equilibrates within a fraction of a sound crossing
time.  The steady state contains large density fluctuations, as in the
free-decay runs.  As in the free decay runs, the velocity and magnetic
field power spectra evolve toward approximately $\propto k^{-2}$ (even
though the model is forced at a scale well below the box size).  The
best fit slope for the combined power spectrum is $s= 2.2\pm 0.12$,
averaged between times $t/\ts= 1$ and $5$.  After saturation, the
density power spectrum is flat below $k = 8 (2\pi/L)$.

We now turn to the ``linewidth-size'' relation in our simulations.  In
molecular clouds, the linewidth-size relation, one of ``Larson's laws''
(\cite{lar81}), is a correlation with the approximate form $\sigma_v(R)
\propto R^{1/2}$ between the linear size $R$ of a region and its
internal velocity dispersion $\sigma_v(R)$.  Any linewidth-size
relation is directly related to the power spectrum of the velocity:
for $|\bv_k|^2 \propto k^{- n - m}$ in $m$ dimensions, we have $\sigma_v(R)
\propto R^{n/2}$ (assuming the emissivity of the gas is uniform).
Larson's law is then a natural consequence of an $n = 1$ power
spectrum.  Power spectra with $n = 1$ occur in systems with an
ensemble of velocity discontinuities, known as Burger's turbulence,
and are observed in the solar wind (\cite{bur87}) and in simulations
of supersonic hydrodynamic turbulence (\cite{pas88},\cite{por92}).
The power spectra in our simulations evolve to approximately
$|v_{\perp,k}|^2,\, |B_{\perp,k}|^2\propto k^{-2}$ (independent of the
forcing scale).  By smoothing with a Gaussian window, we have directly
verified the correspondence between the power spectrum and
linewidth-size relation.  For the standard run we find a
linewidth-size relation of the form $\sigma_v(R) \propto R^{0.57\pm
0.03}$ between scales $L/512$ and $L/2$.

\begin{figure}
\plotone{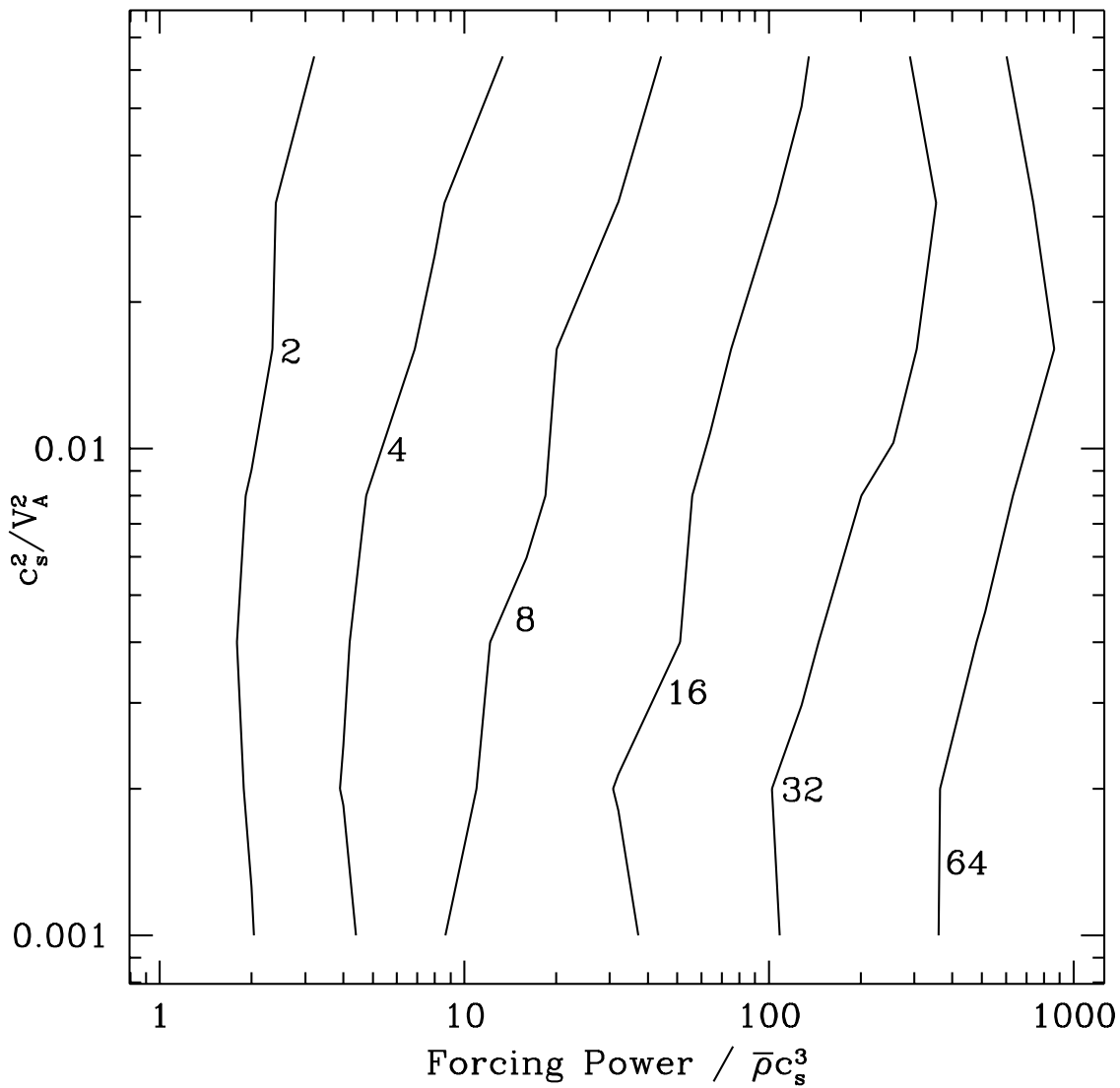}
\caption{
A contour plot of the saturation kinetic energy $E_K$ (defined as the
average
kinetic energy during the final two sound crossing times of the
simulation)
as a function of the forcing power and $\beta \equiv \cs^2/\va^2$.
The contours are located at $E_K/(\rb L\cs^2) = 2, 4, 8, 16, 32, 64$.
}
\end{figure}

We have performed a survey of forced runs to measure the saturation
energy as a function of $\beta$ and the forcing power.  The results are
shown in Figure 8; a fit to the results that is better than a factor of
$2$ everywhere in the region surveyed is
\begin{equation}\label{ESATFIT}
E_W = 0.48 \left({\dot{E}\over{\rb c_s^3}}\right)^{0.67}
	\beta^{-0.16} \rb L c_s^2.
\end{equation}
We can use the results of Figure 8 to decide what input power is
required to sustain a given level of turbulence.  Solving for the
required power based on the fit above gives
\begin{equation}\label{REQPOW}
\dot{E} = 3.0 \left({\Ew\over{\rb L c_s^2}}\right)^{1.5}
	\beta^{0.24} \rb c_s^3.
\end{equation}
Since the model is in equilibrium the input power matches the
dissipation rate.  

Notice that, apart from the factor of $\beta^{0.24} \simeq
(c_s/v_A)^{1/2}$, the energy dissipation rate per unit volume is $\propto \rho
\sigma_v^3/L$, identical to what one would expect for homogeneous,
incompressible turbulence.  This is, in part, just dimensional
analysis: a dissipation rate per unit
volume must have units of a density times a speed cubed over a
length.  But it is not obvious which speed to use in
completing this estimate, since there are three velocity scales in the
problem: the sound speed, the \alf speed, and the velocity dispersion.
The simulations show that the correct speed is a power-law-mean of
these three that is dominated by the velocity dispersion.

An example of an alternative scaling for the \alf-wave dissipation rate
is the estimate of \cite{coh75} for the steepening of a single wave.  The
implied energy dissipation rate per unit area is $\dot \Ew \sim
2\pi\left({\Ew/{\rb L c_s^2}}\right)^{2} \beta^{1/2} \rb c_s^3$.  The
corresponding volumetric energy dissipation rate scales as 
$(\sigma_v/\va )\times(\rb\sigma_v^3/L)$, different from the results of our
simulations.  Thus the scaling of equation (\ref{REQPOW}) is not a
trivial prediction of quasilinear theory, even for compressible
magnetized plasmas.

The dissipation rate in the forced runs depends on the wavenumber at
which the energy is injected ($k_{\rm pk}$).  In the above runs the
energy is injected near $k_{\rm pk} = 8 (2\pi/L)$.  A survey of the
dissipation rate as a function of $k_{\rm pk}$ reveals that the
saturation energy obeys
\begin{equation}
E_W \propto k_{\rm pk}^{-0.30}.
\end{equation}
Together with equation(\ref{ESATFIT}), this implies that
\begin{equation}\label{POWK}
\dot{E} \propto k_{\rm pk}^{0.46},
\end{equation}
so that the dissipation rate increases as the forcing scale decreases.

We can define a wave dissipation time $t_{\rm diss}$ in equilibrium by
taking the ratio of $\Ew$ to $\dot E$ as given in equation
(\ref{REQPOW}). Including the forcing-wavelength dependence of
equation (\ref{POWK}), we find
\begin{equation}\label{TDISS}
t_{\rm diss}\equiv {\Ew\over \dot{\Ew}}= 0.87 \left({\lambda_{\rm
pk}\over L}\right)^{0.46}
\left({\Ew\over{\rb L c_s^2}}\right)^{-1/2}
\beta^{-0.24} {L\over \cs}.
\end{equation}
We can compare the dissipation time for forced disturbances to the
\alf time $\ta$ or, for self-gravitating clouds, to the characteristic
collapse time $\tc$ (eqs. \ref{tcollapse}, \ref{talfven}).  Using
$\Ew/(\rb L \cs^2)\approx 3\sigma_v^2/\cs^2$, the former ratio is
$t_{\rm diss}/\ta \approx 0.5 (\lambda_{\rm pk}/L)^{1/2}(\va/\sigma_v)
(\va/\cs)^{1/2} $; this ratio is generally large unless the forcing
scale is very small.  Similarly, $t_{\rm diss}/\tc\approx 0.5
(\lambda_{\rm pk}/L)^{1/2}(\va/\cs)^{1/2} (\nj \cs/\sigma_v)$; note
that a ratio smaller than unity does not necessitate collapse (see
\S\ref{GRAVDECAY}).

\subsection{Self-Gravitating Decay Runs}\label{GRAVDECAY}

A crucial question for the evolution of molecular clouds is whether the
nonlinear MHD waves interact with the self-gravitating compressive
modes (here generically referred to as ``Jeans modes'') so
as to prevent or delay collapse.  We can examine this issue by
turning on self-gravity in our simulations.

Let us start with a few preliminary words about self-gravity in a
periodic slab geometry.  First, in one-dimensional, slab geometry an
isothermal fluid always has a stable equilibrium, whereas in three
dimensions a self-gravitating, isothermal fluid sphere is not
generally stable (see \cite{spi68}).  In non-periodic slab geometry,
the equilibrium self-gravitating, isothermal density distribution is
\begin{equation}\label{SGSLB}
\rho = \rho_o {\rm sech}^2(z/z_o),
\end{equation}
where $z_o = c_s/\sqrt{2\pi G\rho_o}$ (\cite{spi42}).  In one
dimension, the self-gravitating slab is the nonlinear outcome of the
Jeans instability.  Second, periodicity implies an important change in
the physics of self-gravity: the adoption of the Jeans ``swindle.''
When we solve Poisson's equation we make no allowance for the
self-gravity of the background of material at the mean density.  Thus
we are really solving the equation
\begin{equation}
\nabla^2\phi = 4 \pi G (\rho - \rb)
\end{equation}
rather than Poisson's equation.  

Consider a periodic slab geometry with initially uniform density.  The
self-gravity is such that there are $\nj$ Jeans lengths inside length
$L$. If we assign a binding energy $E_G = 0$ to the state with uniform
density, it is possible to estimate the binding energy of the
self-gravitating slab under the assumption that the density
distribution is very nearly the same as for a non-periodic
self-gravitating slab, equation(\ref{SGSLB}), and that there is only
one slab present inside $L$.  The result is
\begin{equation}\label{EBIND}
E_{G,max} \simeq \left(1 - {1\over{6 n_J^2}} - {n_J^2\pi^2\over{6}}\right) 
	\rb L c_s^2.
\end{equation}
This is the equilibrium with the {\it maximum} possible binding energy
for an isothermal fluid in one dimension.

\begin{figure}
\plotone{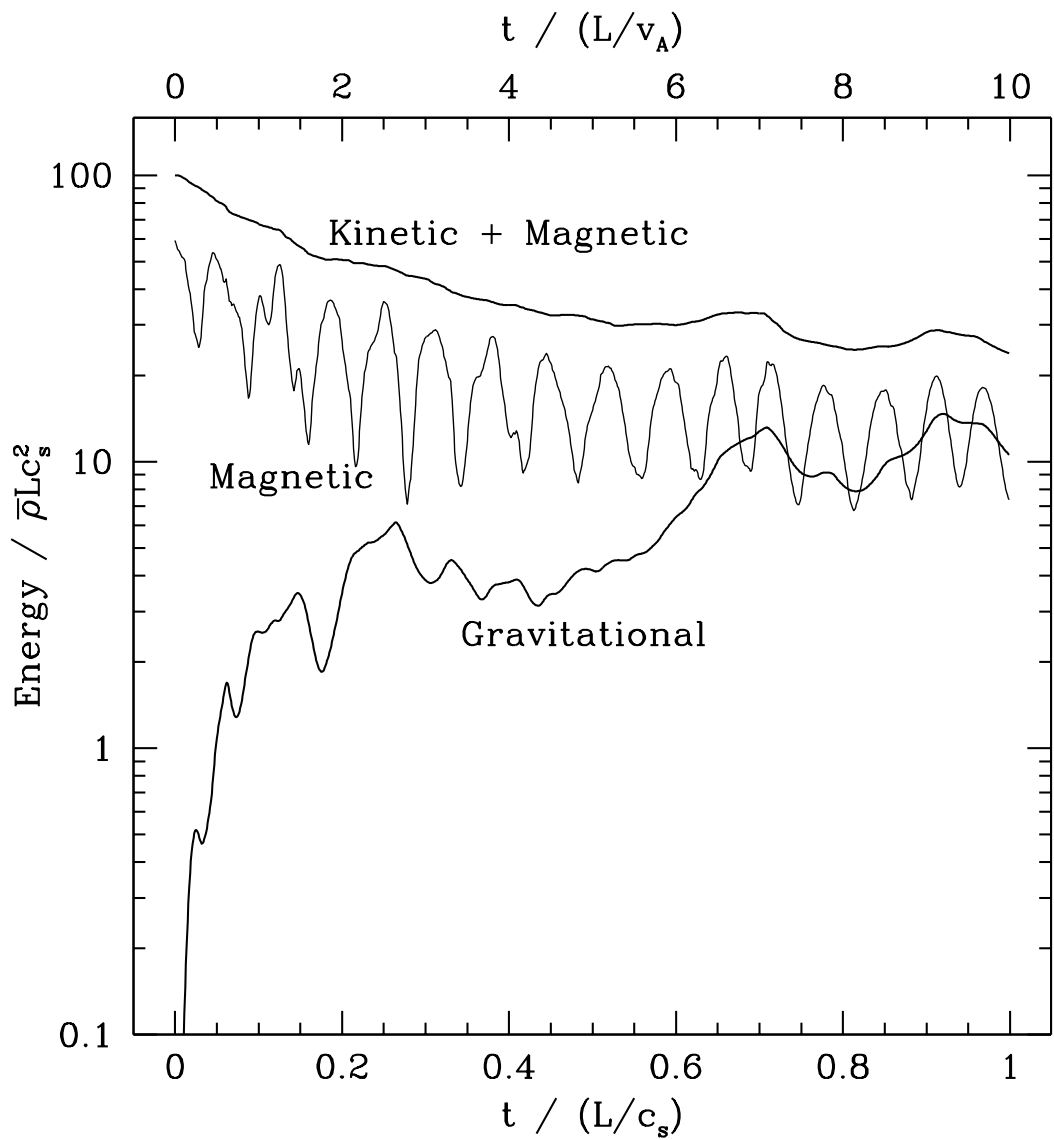}
\caption{
Time evolution of the magnetic+kinetic, magnetic, and gravitational
potential energy in the ``standard'' run with decay and gravity.  There
are four Jeans lengths inside the simulation.
Thus $\tc=0.25\ts=2.5\ta$.
}
\end{figure}

The energy evolution of the standard self-gravitating/decay run, with
$\beta=0.01$, initial wave energy $\Ew=100\rb L c_s^2$, and Jeans
number $\nj=4$ (so that $E_{G,max}\simeq -25 \rb L\cs^2$), is shown in
Figure 9.
\footnote{Since $c_s, \rb,$ and
$L$ are fixed, the number of Jeans length is set by manipulating $G$ so
that, numerically, $G = \pi n_J^2$.} The background of transverse waves
induces large density fluctuations within an \alf crossing time, as in
the non-self-gravitating case.  These density fluctuations now have a
gravitational binding energy, however.  The density peaks grow and
coalesce until at the end of the run (at $t = \ts = 10 \ta = 4 \tc$)
there are only two peaks.  As in the non-self-gravitating case, the
field lines kink inside the density peaks and are nearly straight (i.e.
force-free) in between.  Since $E_G/E_{G,max} \simeq 0.4$ at the end of
the simulation, the background of MHD waves evidently can delay
collapse.

\begin{figure}
\plotone{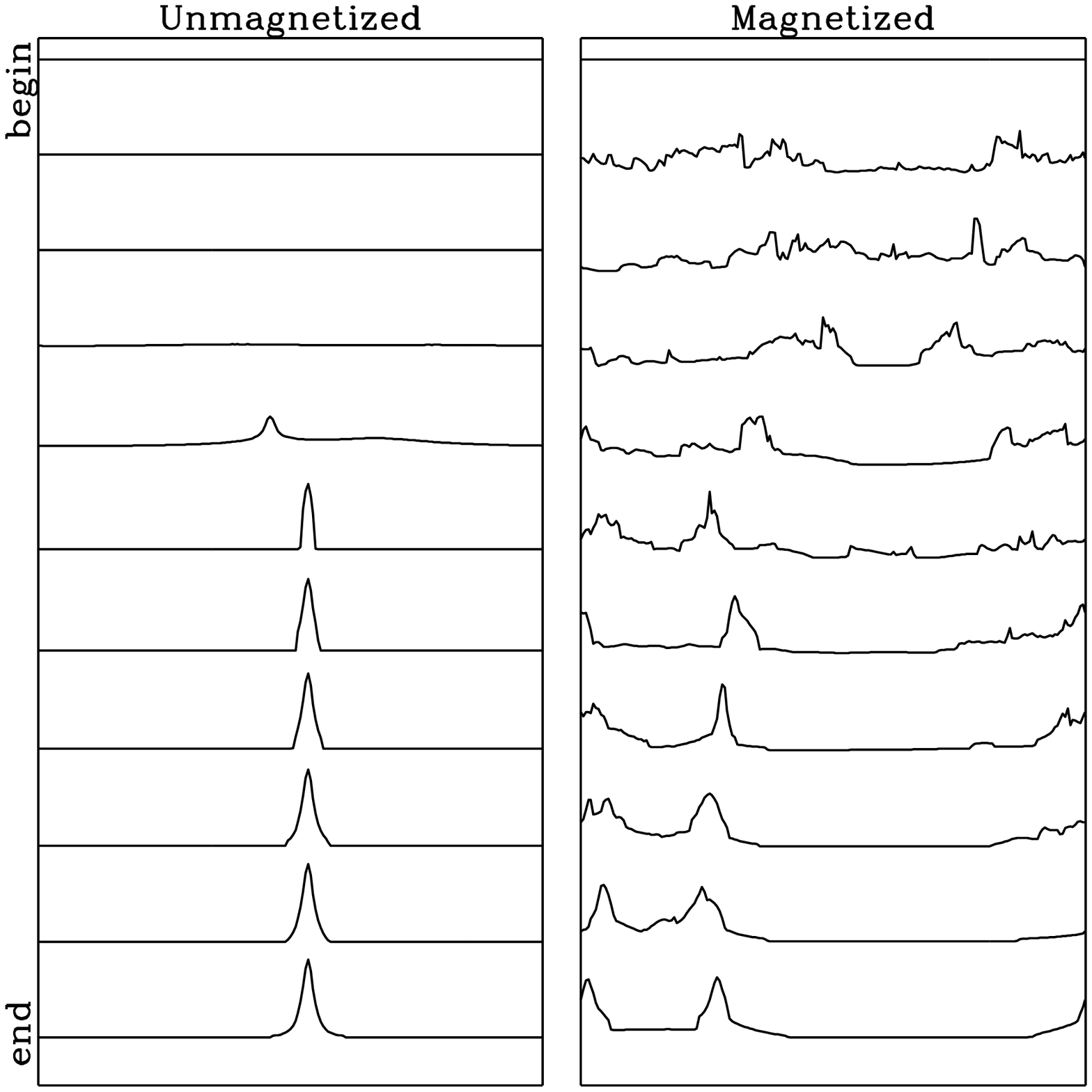}
\caption{
Time evolution of the density structure in the ``standard''
self-gravitating 
run with decaying wave energy (right panel), and in an unmagnetized
control 
run (left panel).  To reduce the contrast, we display $(\rho/\rb)^{1/4}$
for both cases.  Time intervals between the snapshot portraits are
$\Delta t=\ta=0.1\ts=0.4\tc$.  The two dense structures which gradually
coalesce in the magnetic simulation oscillate in the transverse 
direction.
}
\end{figure}

A more detailed look at the evolution of the density structure in the
standard simulation is provided in the right panel of Figure 10, which
shows $(\rho/\rb)^{1/4}$ at intervals of $\ta = 0.1 \ts = 0.4\tc$.  The
density has been raised to the $1/4$ power to reduce the contrast.  The
large density contrast induced by the the initial wave spectrum is
evident at the first snapshot; the density maxima coalesce and grow
over the course of the simulation.  The left panel of Figure 10 shows a
nonmagnetic simulation that begins with low amplitude white noise in
$\rho$ and $v_x$.  The nonmagnetic simulation provides a control for
the magnetic simulation and shows that the presence of transverse MHD
waves broaden the existing density maxima and delay or prevent their
coalescence over several collapse times.

How are the density and magnetic pressure correlated?  As in \S
\ref{FREEDECAY}, we have smoothed the density and magnetic pressure on a
scale $\lambda$, taken the logarithm of each, and measured the slope
and strength of the correlation over a range of $\lambda$.  When
smoothed on the smallest scales, the density and magnetic pressure are
weakly correlated (averaging over time), with $\delta B^2 \propto
\rho^{0.1}$.  As $\lambda$ is increased, however, the strength and
slope of the correlation increase.  When $\lambda = L/5$ the slope and
strength of the correlation have approximately doubled.  Thus on large
scales the Jeans modes and \alf waves do tend to interact in such a way
as to resist collapse.  Because the correlation is weak and
variable, however, it is inappropriate to interpret it as an
effective polytropic equation of state for the wave pressure.

\begin{figure}
\plotone{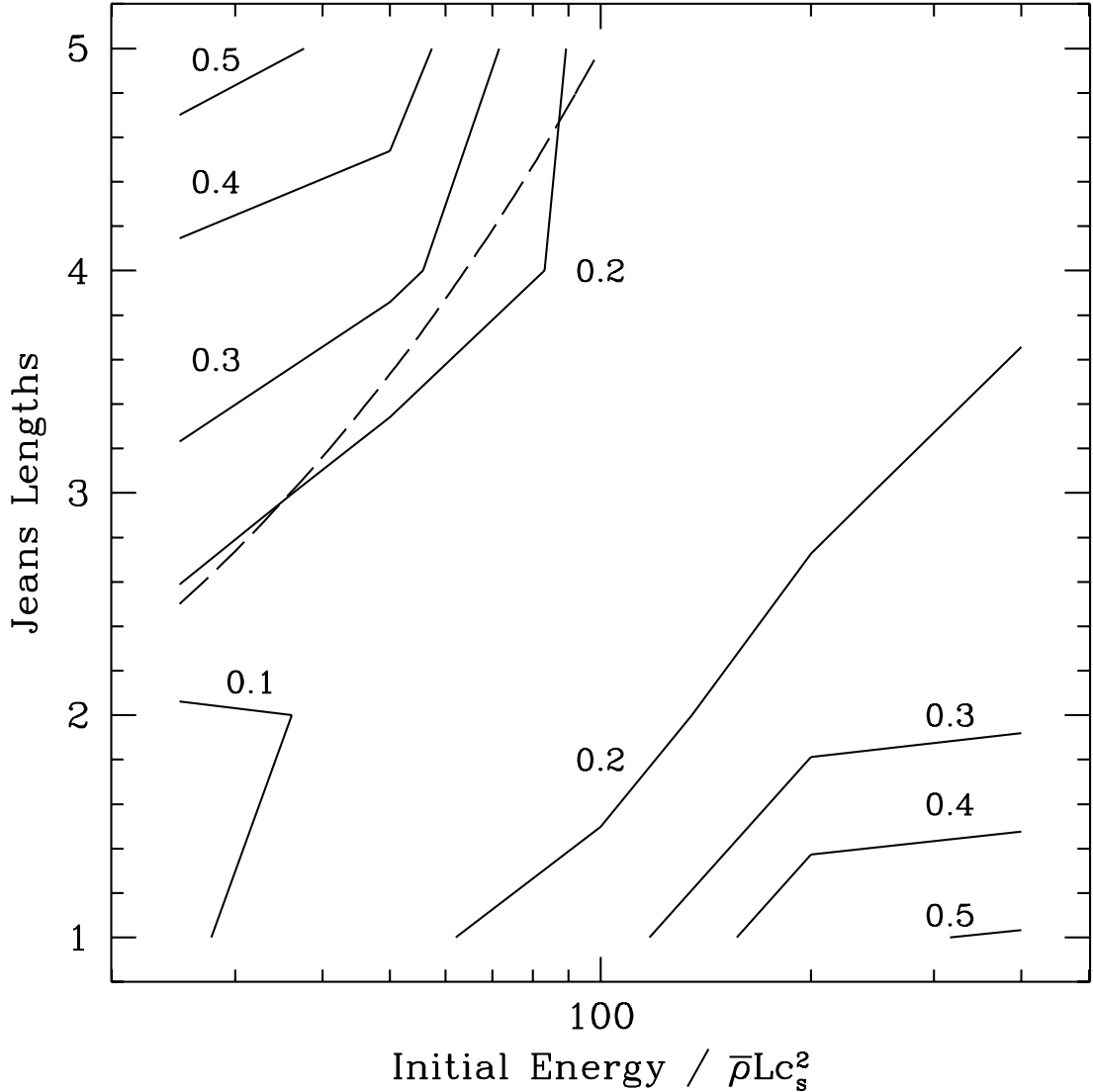}
\caption{
A contour plot of the gravitational potential energy $E_{\rm G}$
relative to the maximum gravitational potential energy for a
fully-collapsed sheet $E_{\rm G,\, max}$ (eq. 26), averaged between 0.9
and 1.5 $\tc$ (the collapse time $\tc=\ts/\nj=10\ta/\nj$ for
$\beta=0.01$, where $\nj=L/\Lj$ is the number of Jeans lengths in the
simulation) The ratio $E_{\rm G}/E_{\rm G,\, max}$ is measured as a
function of the initial magnetic+kinetic energy and $\nj$.  A large
value of $E_{\rm G}/E_{\rm G,\, max}$ corresponds to strong density
concentration.  The dashed curve shows the locus $\nj=(E_{\rm W,\,
init}/\rb L\cs^2)^{1/2}/2$ (eq. [27]) predicted by the pseudo-Jeans
analysis (\S 2.2) as the lower boundary of the collapsed region.  See
text for a discussion of the low-$\nj$, high-$\Ew$ part of the
diagram.
}
\end{figure}

When can unforced MHD waves prevent gravitational collapse?  To
investigate this question, we have performed simulations with different
initial energies and number of Jeans lengths.  We measure the degree of
collapse by comparing the time-averaged (from $0.9 \tc$ to 1.5 $\tc$)
gravitational binding energy to the gravitational binding energy for a
fully collapsed sheet, given by equation (\ref{EBIND}).  Figure 11
shows contours of this ratio ($E_G/E_{G,max}$) as a function of the
initial wave energy and the number of Jeans lengths in the simulation.

The maximum in $E_G/E_{G,max}$ in the upper left corner of Figure 11 is
where collapse has occurred.  This is qualitatively consistent with the
pseudo-Jeans analysis of \S \ref{S: cloud properties}, which suggests
that collapse should occur when the gravitational energy 
overwhelms the wave energy.  Perhaps coincidentally, the pseudo-Jeans 
analysis
is nearly quantitatively correct as well.  The dashed line in the Figure shows
the locus
\begin{equation}\label{NJMIN}
\nj= {1\over 2}\sqrt{E_{\rm W,\, init}\over\rb L \cs^2},
\end{equation}
predicted by equation (\ref{PSJEANS}), when $\Ew \gg \rb L \cs^2$, as
the lower limit in Jeans number for collapse to occur.  This locus is
indeed close to the boundary of the ``collapsed'' region.  Other
simulations show that the structure of this diagram is only weakly
dependent on $\beta$ over the range we have surveyed.

There is also a maximum in $E_G/E_{G,max}$ in the lower right corner of
Figure 11.  In this region of parameter space the large initial wave
motions drive large density fluctuations.  These density fluctuations
have an associated binding energy, which is large compared to the
maximum possible binding energy $E_{G,max}$ because $E_{G,max}$ is
small when $n_J$ is small.  Thus, while the binding energy is near
maximal, the system is not bound because the wave energy is
comparatively large.

The scaling with energy of the stability criterion (equation
[\ref{NJMIN}]) and the dissipation timescale (equation [\ref{TDISS}])
leads to the following peculiar situation:  in the {\it stable} region
at the lower right of Figure 11, the dissipation time is {\it
shorter} than the collapse timescale, while in the {\it unstable}
region  in the upper left of the Figure the dissipation time is
{\it longer} than the collapse timescale.  This apparent paradox is
resolved when one realizes that stability is determined by the energy
content of the cloud and not its energy loss rate.

\subsection{Self-Gravitating Forced Runs}\label{GRAVFORCED}

A final set of simulations considers stochastically forced runs with
self-gravity.  These runs are forced in the same manner as the
nonself-gravitating, forced runs of \S\ref{SUBSFEQR},
with the peak forcing at $k_{\rm pk} = 8\,(2\pi/L)$.  The
simulations are run without self-gravity for $0.5$ sound crossing
times, so that the model has a chance to equilibrate; then gravity is
turned on.

\begin{figure}
\plotone{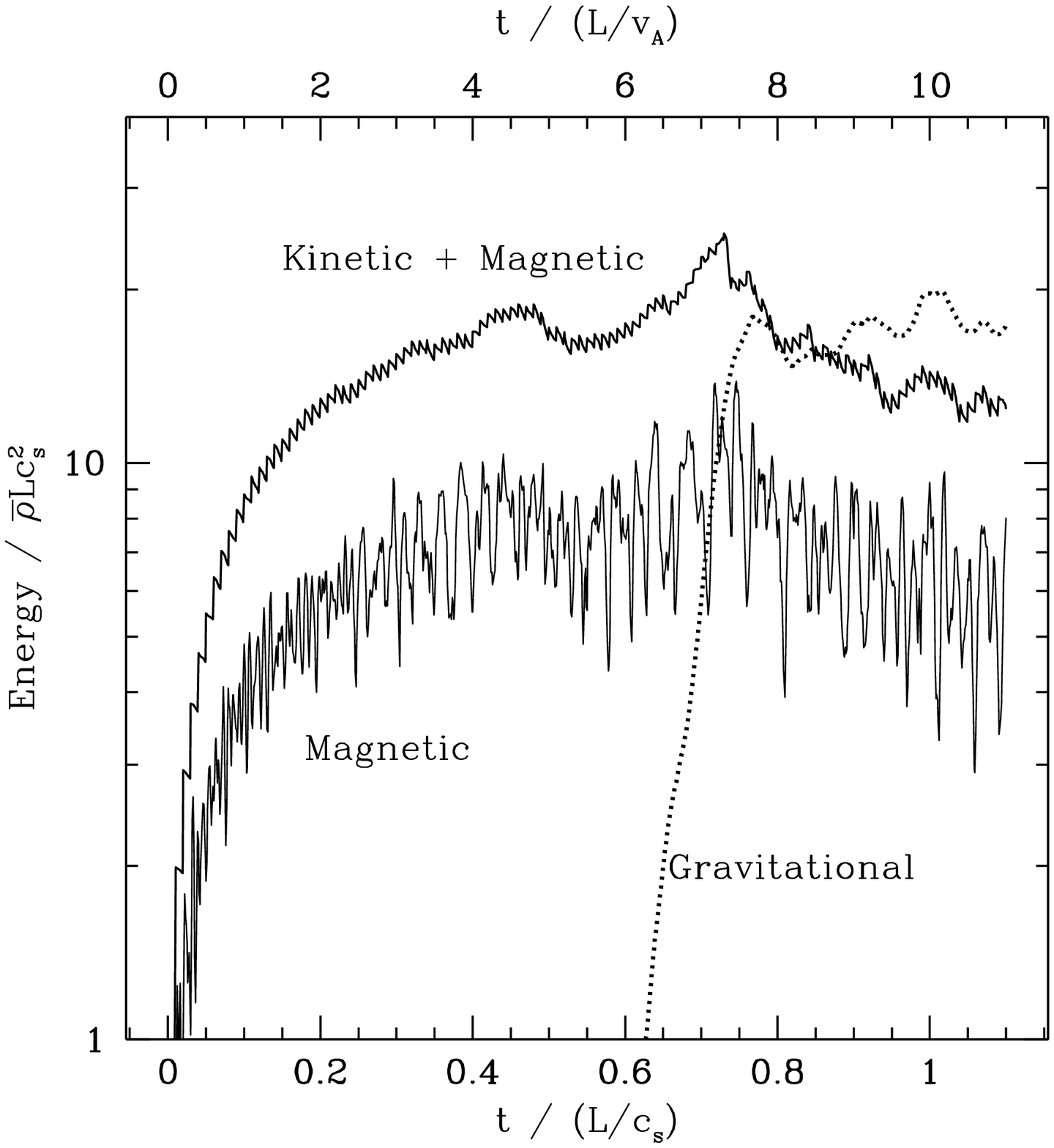}
\caption{
Time evolution of the magnetic+kinetic, magnetic, and gravitational
potential energy in the ``standard'' run with forcing and gravity.  The
strength of gravity is such that there are four Jeans lengths inside the
simulation scale, and the forcing power is $100 \rb c_s^3$.
}
\end{figure}

The standard run, with $\beta=0.01$, $\nj=4$ 
(so $E_{G,max}\simeq -25 \rb L\cs^2$), and forcing power $\delta
E/\delta t = 100 \rb c_s^3$, is shown in Figure 12.  The initial $0.5$
sound crossing times is identical to the nonself-gravitating runs;
collapse occurs as soon as self-gravity is turned on.  The longitudinal
motions associated with collapse cause a peak in kinetic energy near
this time.

\begin{figure}
\plotone{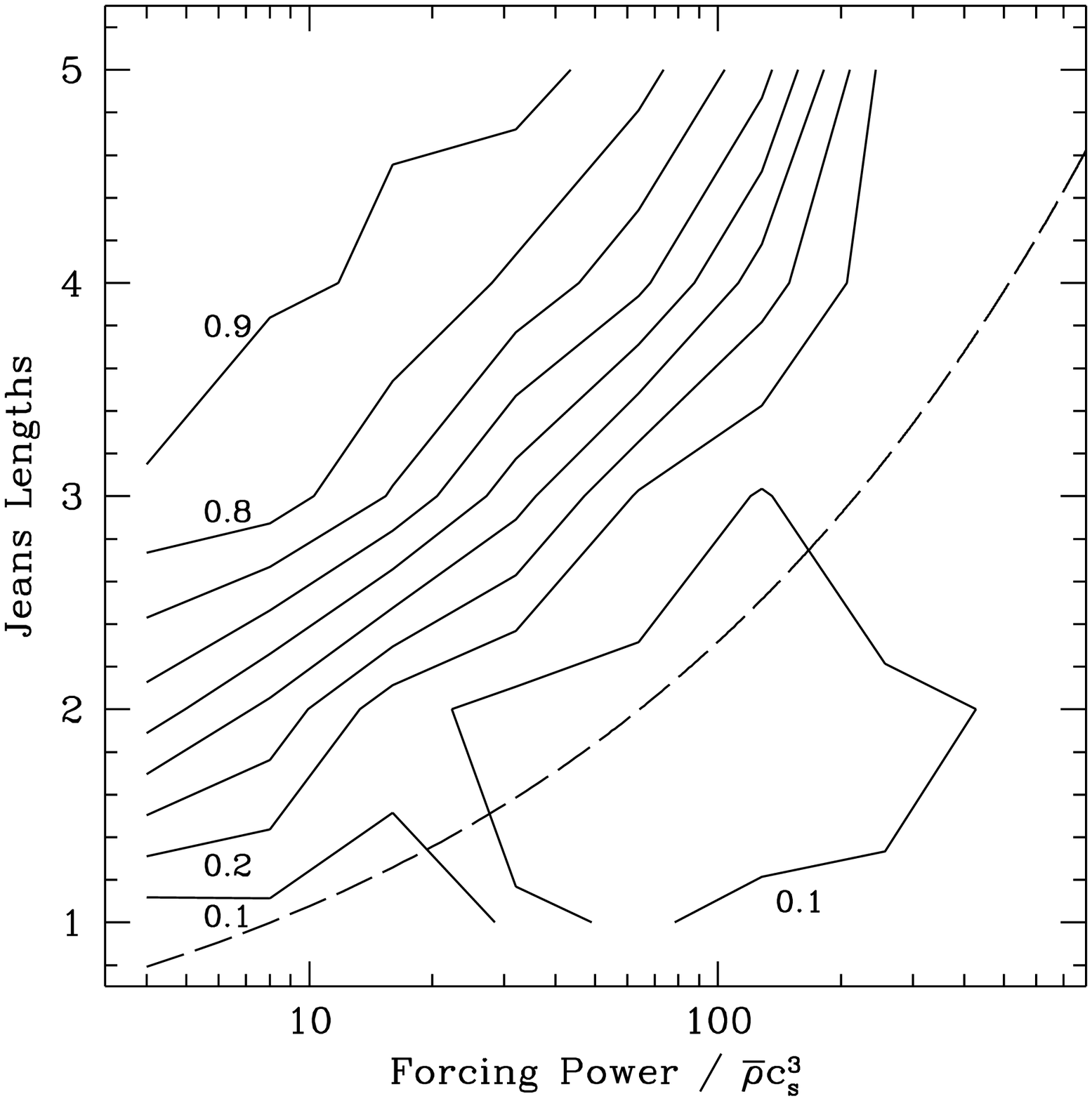}
\caption{
A contour plot of the gravitational potential energy $E_{\rm G}$
relative to the maximum gravitational potential energy for a
fully-collapsed sheet $E_{\rm G,\, max}$ (eq. [26]) measured at the end
of the forced, self-gravitating simulations.  $E_{\rm G}/E_{\rm G,\,
max}$ is found as a function of the forcing power and Jeans number
$\nj=L/\Lj$.  The dashed curve shows the locus estimated as the lower
limit for collapse in equation (28).
}
\end{figure}

Can collapse be prevented in these forced runs?  We have performed
simulations for a variety of $n_J$ and forcing powers and the results
are displayed in Figure 13.  The contours in Figure 13 lie at constant
$E_G/E_{G,max}$.  The pseudo-Jeans analysis leading to equation
(\ref{NJMIN}), together with the saturation energy predicted by the
fit to the nonself-gravitating, forced runs (equation [\ref{ESATFIT}];
the self-gravitating runs have a saturation wave energy that is comparable
with this result, but generally smaller by a factor of $< 2$),
predicts that collapse should occur for
\begin{equation}\label{NJMINF}
\nj > 0.35 \beta^{-0.08}\left({\dot{E}\over \rb \cs^3}\right)^{0.33}.
\end{equation}
The predicted locus of marginal stability for $\beta = 0.01$ is shown
as a dashed line in Figure 13.  The weak $\beta$ dependence in equation
(\ref{NJMINF}) has been confirmed by runs not shown in the figure.  The
pseudo-Jeans analysis is thus a fair collapse predictor in this case as
well.  The corresponding lower limit for the power input needed to
prevent collapse is
\begin{equation}\label{REQPOWG}
\dot E = 24 \beta^{0.24} \nj^3 \rb\cs^3,
\end{equation}
which for $\beta=0.01$ has a coefficient $7.9$.  In any event, it is
clear that transverse magnetic pressure can also prevent collapse in
the forced runs.

\section{Discussion}\label{DISCUSS}

\subsection{Applications}\label{APP}

In this section we translate our results into an astronomical context
by applying them to a ``typical'' molecular cloud.  This is a 
speculative venture, since our simulations employ slab symmetry
and make a number of other drastic approximations.  The exercise seems
worthwhile, however, since the simulations are self-consistent and
fully nonlinear, unlike earlier treatments of the problem.  Are our
results in harmony with the known, observed properties of molecular
clouds?

Consider a representative molecular cloud of mean linear dimension $L=
20 \pc$, number density $100 \cm^{-3}$, kinetic temperature $T = 20
\K$, and line-of-sight velocity dispersion $\sigma_v = 2 \kms$.  For
mean field strength $B_0 = 20 \mu\G$, $\beta = 0.01$.  The
sound-crossing, gravitational collapse, and \alf-crossing timescales
for this reference cloud are $\ts = 74\Myr$, $\tc = 9.9 \Myr$, and
$\ta=7.6 \Myr$ (assuming solar metallicity).  The Jeans number $\nj =
L/\Lj = \ts/\tc = 7.5$, and the ratio $\ta/\tc = 0.76$.  Thus, for a
static uniform field, the cloud would be unstable parallel and stable
perpendicular to the mean field (see \S \ref{S: equilibrium}).  If the 
projected area of the cloud is
$L^2 = 400 \pc^2$, then the kinetic energy per unit surface area is
$(3/2)\rb\sigma_v^2 L = 87 \rb L \cs^2$.  Assuming equipartition
between kinetic and perturbed magnetic energy, the wave energy per unit
surface area is twice the kinetic energy, $E_W = 173 \rb L c_s^2$.
Finally, if the cloud's volume is $L^3$, its total mass is $5.6 \times
10^4 \msun$.

In the absence of self-gravity or any energy inputs, the cloud would,
by assumption, evolve like the free decay simulations of \S
\ref{FREEDECAY}.  In particular, it should evolve similarly to the
``standard'' decay of Fig. 2, which has $E_W = 100 \rb L c_s^2$ and
$\beta=0.01$, and is shown for comparison to the other decays as the
central dotted curve in Fig. 5.  To provide a more precise comparison,
we have run a decay simulation starting from $E_W = 173 \rb L c_s^2$.
After $9.6\Myr (=0.16\ts)$, the wave energy would drop by factor
$\approx 2$, so the velocity dispersion would drop by a factor of
$\sqrt{2}$; after $41\Myr$ by a factor $\approx 4$, corresponding to a
factor of $2$ drop in the velocity dispersion.  At $15\Myr$, the cloud
contains clumps.  Fully $28 \%$ of the mass, but only $2 \%$ of the
volume, lies at densities greater than $10^3 \cm^{-3}=10\rb$, while
$74 \%$ of the volume lies at densities below the mean.

Now suppose the cloud's internal motions are forced (by winds from
young stars, for example) at a scale $\lambda = L/8 = 2.5
(L/20\pc)\pc$, such that the mechanical power input $\cal P$ equals the
wave dissipation rate $\cal L$ and the observed internal wave energy
represents a saturated equilibrium.  Again, we temporarily ignore
self-gravity.  We can use the forced, non-self-gravitating simulations
to estimate the required input power and dissipation rate associated
with a given equilibrium level of internal kinetic energy.  In general
the dissipation rate $\cal L$ should be roughly equal to the luminosity
of the cloud in the important gas phase cooling lines, while the
mechanical power input $\cal P$ would equal the sum of energy inputs
from young stellar winds (\cite{nor80}), \alf waves emerging from
collapsing, rotating cloud cores (\cite{gil79a}, \cite{gil79b},
\cite{mou80}), etc.  In equilibrium ${\cal P}={\cal L}$.  Using
equations (\ref{REQPOW}) and (\ref{POWK}), we find 
\begin{equation}\label{INPOW}
{\cal P}={\cal L}= 19
	\left({\sigma_v\over{2 \kms}}\right)^3
	\left({n_{H_2}\over{100 \cm^{-3}}}\right)
	\left({L\over{20\pc}}\right)^2
	\left({\beta\over{0.01}}\right)^{1/4}
	\left({\lambda\over{2.5\pc}}\right)^{-1/2}
	\lsun.
\end{equation}
A dissipation timescale can be estimated via $t_{diss} = E/\dot{E}$
(eq. \ref{TDISS});
the result is
\begin{equation}\label{DISST}
t_{diss} = 6.0 
	\left({\sigma_v\over{3 \kms}}\right)^{-1}
	\left({L\over{20\pc}}\right)
	\left({\beta\over{0.01}}\right)^{-1/4}
	\left({\lambda\over{2.5\pc}}\right)^{1/2}
	\Myr.
\end{equation}
This dissipation timescale is comparable to the collapse timescale.

In fact, our reference cloud {\it is} self-gravitating; it contains $n_J^3 = 
(7.5)^3 = 410$ Jeans masses.  While computational expense prevents us 
from performing simulations with $n_J \gtrsim 5$, we
find that equation (\ref{NJMIN}) is a good predictor of 
collapse on a dynamical timescale when $n_J \le 5$.  
Assuming equation (\ref{NJMIN}) applies for more strongly 
self-gravitating clouds as well, we find that immediate contraction can be 
avoided when
\begin{equation}\label{BNOCOLL}
\<|\delta \bB|^2\>^{1/2} > 18 
	\left({L\over{20 \pc}}\right)
	\left({n_{H_2}\over{100 \cm^{-3}}}\right)
	\mu\G,
\end{equation}
corresponding to 
\begin{equation}\label{NOCOLL}
\sigma_v > 2.3 \left({L\over{20 \pc}}\right)
	\left({n_{H_2}\over{100 \cm^{-3}}}\right)^{1/2}
	 \kms.
\end{equation}
Our reference cloud with $\sigma_v =2\kms$ has velocity slightly below
the threshold level, and in the absence of energy inputs
would contract on a timescale $\sim \tc$.

The reference cloud could be supported in an indefinite equilibrium if
it were supplied with energy at a scale $\lambda$, as in the forced,
self-gravitating simulations.  Using equations (\ref{REQPOWG}) and 
(\ref{POWK}), we
find that an input power of
\begin{equation}\label{INPOWG}
{\cal P}= 27 
	\left({L\over{20 \pc}}\right)^{5}
	\left({n_{H_2}\over{100 \cm^{-3}}}\right)^{5/2}
	\left({\beta\over{0.01}}\right)^{1/4}
	\left({\lambda\over{2.5\pc}}\right)^{-1/2}
	\lsun
\end{equation}
or greater is required to support the cloud in a true equilibrium.  A smaller
supply of power would lead to gradual contraction.  Notice that the
power required to support the current level of turbulent energy in our
reference cloud (eq. \ref{INPOW}) would be insufficient to support it 
indefinitely against gravity.  For the latter, the higher power level 
of equation (\ref{INPOWG}) would be required, and the one-dimensional
velocity dispersion would be raised to the  level indicated in
equation (\ref{NOCOLL}).

The ``reference'' cloud described above is meant to represent a cloud
like Orion A (\cite{bal87}) or the Rosette nebula (\cite{wil95}).  For
clouds as large as this, we have had to extrapolate our fits to larger
values of the Jeans number and input power than we were able to
compute directly.  On the other hand, smaller self-gravitating clouds
such as Taurus-Auriga (\cite{cer91}) or Ophiuchus (\cite{lor89a},
\cite{lor89b}) are within the parameter range we have simulated, and
therefore the results presented in the figures of \S 4 can be used
directly.  Diffuse, high-latitude clouds may be best represented by our
unforced, non-self-gravitating simulations.

The power requirements for sustaining turbulence and counteracting
gravity of equations (\ref{INPOW}) and (\ref{INPOWG}) are reasonable
for GMCs.  For example, the total hydrodynamic outflow momentum
observed in Orion A (which has properties  comparable to our
reference cloud) is estimated at $320 \msun \kms$ (\cite{fuk93}),
which for mean outflow velocity $30\kms$ and outflow lobe size
$0.75\pc$ (cf. \cite{fuk89}) implies a characteristic power input of
$33\lsun$.  Thus our results are in accord with
observations.

Finally, let us suppose that clouds evolve in quasi-equilibrium, and
support can ultimately be ascribed to the power originating in young
stellar outflows.  Then by equating the required power ${\cal P}$ to
the total wind mechanical luminosity $G \dot{\cal M}_* M_*/R_c$ we can
obtain a total star formation rate $\dot{\cal M}$ in the cloud.  Here
we absorb the details of the wind acceleration mechanism 
into the characteristic radius $R_c$ where the wind originates (see,
e.g.,
\cite{shu94}), and the uncertainties associated with the coupling of
the wind to the rest of the cloud into the scale $\lambda$.  Scaling
$R_c$ to $0.1 \au$ and using equation (\ref{INPOWG}), we find
\begin{equation}
\dot{\cal M}_* = 2 \times 10^{-5}
	\left({L\over{20 \pc}} \right)^5
	\left({n_{H_2}\over{100 \cm^{-3}}} \right)^{5/2}
	\left({\beta\over{0.01}} \right)^{1/4}
	\left({\lambda\over{2.5\pc}} \right)^{-1/2}
	\left({R_c\over{0.1 \au}} \right)
	\left({M_*\over{\msun}} \right)^{-1}
	\msun \yr^{-1}.
\end{equation}
This implies a star formation timescale $M(cloud)/\dot{\cal M}_* \simeq
3 \times 10^9 \yr$, which is about $300$ times the dynamical collapse time.
It also implies that over a cloud lifetime of perhaps $3 \times 10^7 \yr$
the cloud will turn about $1\%$ of its mass into stars.

\subsection{Perspective} \label{PERSP}

It behooves us to remind the reader of some of the shortcomings of our
treatment.  The most severe is the use of slab (one dimensional)
symmetry.  In higher dimensions, dissipation rates may
rise because new, transverse decay modes will be available.  On the
other hand, dissipation rates could fall somewhat because clumps, once 
formed, need not collide with each other, as they do in one dimension.
Fortunately, an extension to two dimensions is immediately
practicable.

Although the use of slab symmetry is likely the leading source of error
in our calculation, we have made other approximations that contribute
as well, and that could be relaxed in future treatments of this
problem.  In this work, we have used an isothermal equation of
state.  Numerical work has shown, however, that realistic molecular
cooling can have a significant effect on fragmentation during
gravitational collapse (e.g. \cite{mon91}).  We also neglect ambipolar
diffusion.  This is likely to have particularly significant effects at small 
scales
in clouds, where MHD waves cannot propagate.  Even for \alf waves on
the scale of the GMCs, the ambipolar-diffusion damping timescale can be
smaller than the dynamical timescale if the only source of
ionization is cosmic rays at a rate $10^{-17} s^{-1}$.  Overall, inclusion
of ambipolar diffusion would tend to increase total dissipation rates, and
to steepen the power spectrum.

Another idealization in our simulations is the periodic boundary
conditions.  These boundaries prevent any wave energy losses from our
model clouds by radiation to an external medium, as would reflecting
boundary conditions.  Estimates indicate that under certain conditions
wave radiation may dominate other energy losses from molecular clouds (\cf
\cite{elm85}).  For a linear amplitude wavetrain, the transmission
coefficient at the cloud edge is small when the density changes sharply
over a distance smaller than the wavelength.  Since most of the energy in 
our simulations is at the largest scales, our results could be sensibly 
applied to clouds with edge gradients sharper than the inverse of
the cloud size.

Finally, we have neglected cosmic ray pressure and transport,
radiative transfer effects, and, of course, feedback from star
formation.  It is not likely that these effects can be incorporated in
any realistic way in the near future.  Our forcing algorithm does
model power input by, e.g., stellar winds, but only in a crude
fashion.

Nevertheless, our treatment does represent significant progress.  As
far as we are aware, it is the first fully self-consistent treatment of
a turbulent, magnetically dominated, compressible, self-gravitating
fluid.  Our self-consistent simulations with realistic field strengths
and velocities lead to a highly inhomogeneous state characterized by
MHD discontinuities.  This state bears
little resemblance to any regime that has been well studied in the
past.  In particular, while some results carry over from quasi-linear
theories and theories of incompressible, eddy-dominated turbulence,
neither can adequately represent the internal dynamics of molecular
clouds.

\subsection{Summary} 

In this paper, we set out to explore the hypothesis that magnetic
forces are crucial to the internal dynamics of Galactic molecular
clouds.  We were motivated by the proposal that the fluctuating
velocity field in MHD waves is responsible for the observed hypersonic
turbulence in molecular clouds, while the associated magnetic field
fluctuations provide a pressure vital in supporting clouds against
gravitational collapse and confining non-self-gravitating clumps.  In 
particular, we were interested in whether
observed molecular clouds represent dynamical equilibria, and more
generally what is required to sustain a given level of
MHD turbulence in the face of nonlinear dissipation and
self-gravity.  A quasistatic equilibrium is possible even with wave
decay;  cloud turbulence may be replenished for example by the gravitational
potential energy liberated from rotating cloud cores and disks when
they collapse and accrete to form stars.

This work uses numerical simulations to investigate large-amplitude
($\delta v/\va\sim 1$) MHD turbulence under density, temperature, and
magnetic field conditions appropriate for Galactic molecular clouds.
Computational expense has imposed some sacrifices of realism in this
first self-consistent treatment of turbulence in a highly compressible,
magnetically-dominated  ($\cs/\va \ll 1$) fluid.  Our main idealizations
are to restrict the dynamics to plane-parallel geometry with periodic
boundary conditions, and to ignore ion-neutral slippage.  This is the
simplest possible model that incorporates magnetic fields and gravity
self-consistently.

We have performed four types of numerical simulations.  In the first
(\S\ref{FREEDECAY}), we evolved a spectrum of
large-amplitude \alf-wave turbulence to determine the free decay rate.
In these simulations, we also investigated the spectral evolution of
the turbulence and formation of density structure.  In our second set
of simulations (\S\ref{SUBSFEQR}), we set out to evaluate the level of
internal stochastic forcing (intended to model power inputs like young
stellar winds) required to sustain a given level of turbulent kinetic
energy.  In our third set of simulations (\S\ref{GRAVDECAY}), we
included self-gravity in model clouds initiated with differing levels
of wave energy to establish collapse thresholds while allowing for
turbulent decay.  In our final set of simulations (\S\ref{GRAVFORCED}),
we applied stochastic internal forcing to self-gravitating clouds to
evaluate the power input needed to prevent collapse.  As a
template for using the dimensionless results of \S\ref{S: Simulations},
we translate into physical units for a ``reference'' cloud in
\S\ref{APP}.

Our major conclusions are as follows: (1) We have confirmed that
nonlinear disturbances in ideal MHD can support a model cloud against
gravitational collapse, provided that the perturbed magnetic energy is
maintained at a level exceeding the binding energy of the cloud (cf.
eq.[\ref{BNOCOLL}]).  Gravity is opposed by a gradient in the
time-varying magnetic pressure due to the components of the field
perpendicular to the mean field.  (2) We have characterized the
dynamical state of highly nonlinear, $\fbeta \ll 1$ MHD systems.  Such
systems contain strong density contrasts, with much of the volume
effectively evacuated and most of the mass concentrated into small
regions.  High density ``clumps'' form and disperse over time, with a
secular trend toward increasingly large concentrations and coherent
motions as energy cascades to larger scales.  These nonlinear,
magnetically dominated systems contain numerous MHD discontinuities,
which naturally give rise to a wave energy power spectrum
approximately $|B_{\perp,\,k}|^2\propto |v_{\perp,\,k}|^2
\propto k^{-s}$ with $s\sim 2$, or linewidth-size relation approximately 
$\sigma_v(R) \propto
R^{1/2}$.  The shape of the power spectra at late times are
essentially independent of the initial input spectrum or energy
injection scale.  (3) We have calculated a decay rate for nonlinear
MHD waves in slab symmetry by equating forcing and dissipation rates
in a saturated equilibrium.  The dissipation time, given in physical
units in equation (\ref{DISST}), is longer than some naive estimates.
In particular, when the cloud is stirred at scales comparable to its
size, the dissipation time exceeds the ``eddy turnover time''
$L/\sigma_v$ by $(\va/\cs)^{1/2}$ times an order-unity factor, and the
\alf-wave crossing time by an additional factor $\va/\sigma_v$ (see
eq.[\ref{TDISS}]).  Peak dissipation rates for free decay obey
approximately the same scaling in $\sigma_v$, $\va$, and $\cs$ as the
dissipation rate in saturated equilibrium.  Because of the present
restricted geometry and negligible friction, our computed
dissipation rates are likely lower limits.

\acknowledgements

We are especially grateful to Jim Stone for helping to 
initiate this project.  We would also like to thank 
Bruce Elmegreen,
Chris McKee,
Phil Myers, 
and
Phil Solomon,
for interesting discussions, 
Alyssa Goodman, 
Jeremy Goodman, 
and
Jerry Ostriker for
comments on a draft of this paper,
and the referee
Ellen Zweibel 
for a discerning review.
This work was supported in part by NASA grant NAG 52837.

\clearpage

\end{document}